\documentclass[twocolumn]{aastex701}
\usepackage{threeparttable}

\setcitestyle{citesep={,}}
\usepackage{hyperref}
\usepackage{amsmath}

\shorttitle{Ice Mantles in YSOs with SPHEREx}
\shortauthors{Serna et al.}

\graphicspath{{./}{figures/}}

\begin{document}

\title{Evolutionary Demographics of Molecular Ice Mantles in Young Stars with SPHEREx}

\correspondingauthor{Javier Serna}
\email{jserna@ou.edu}

\author[0000-0001-7351-6540]{Javier Serna}
\affil{Homer L. Dodge Department of Physics and Astronomy, University of Oklahoma, Norman, OK 73019, USA}
\email{jserna@ou.edu}

\author[0000-0001-9590-2274]{Sean P. Matt}
\affiliation{Homer L. Dodge Department of Physics and Astronomy, University of Oklahoma, Norman, OK 73019, USA}
\email{sean@ou.edu}

\author[0000-0003-1878-327X]{Melissa McClure}
\affil{Leiden Observatory, Leiden University, PO Box 9513, 2300 RA Leiden, The Netherlands}
\email{mcclure@strw.leidenuniv.nl}

\author[0000-0002-5365-1267]{Marina Kounkel}
\affil{Department of Physics and Astronomy, University of North Florida, 1 UNF Dr., Jacksonville, FL 32224, USA}
\email{marina.kounkel@unf.edu}

\author[0000-0001-8600-4798]{Carlos G. Román-Zúñiga}
\affil{Instituto de Astronom\'{i}a, Universidad Nacional Aut\'{o}noma de M\'{e}xico, Ensenada, B.C, M\'{e}xico}
\email{croman@astro.unam.mx}

\begin{abstract}
\noindent Molecular ices are a primary reservoir of volatile C, O, and N in star-forming environments, tracing the chemical pathway from molecular clouds to planet-forming disks. SPHEREx provides contiguous 0.75--5~$\micron$ all-sky spectrophotometry for vastly larger samples than are accessible to JWST, enabling population-scale ice demographics for the first time. We present a search for ice signatures in SPHEREx spectra of 20{,}292 young stellar objects (age $<20$~Myr) drawn from the Gaia-based kinematic catalog of Kounkel \& Covey 2019.
Using laboratory templates for the spectral signatures of H$_2$O (3.05~$\micron$), CO$_2$ (4.27~$\micron$), and CO (4.67~$\micron$), we identify 746 H$_2$O ice candidates, including 131 high-confidence H$_2$O+CO$_2$ detections and 22 with additional corroborating CO absorption.
When binning by age, we find that the H$_2$O$+$CO$_2$ detection rate declines by an order of magnitude between $\lesssim$3~Myr and $\sim$5~Myr, suggesting a chemical or physical evolutionary trend. Candidates concentrate toward high-extinction sightlines ($A_G \gtrsim 2.5$~mag) and embedded SED morphologies, consistent with the idea of envelope dispersal and photodesorption depleting detectable ice mantles within the first few million years of pre-main-sequence evolution. This catalog provides a homogeneous demographic baseline and a prioritized target list for JWST follow-up.
\end{abstract}

\keywords{
\href{http://astrothesaurus.org/uat/1834}{Young stellar objects (1834)} --
\href{http://astrothesaurus.org/uat/2285}{Infrared spectroscopy (2285)} --
\href{http://astrothesaurus.org/uat/849}{Interstellar molecules (849)} --
\href{http://astrothesaurus.org/uat/2092}{Ice Formation (2092)} --
\href{http://astrothesaurus.org/uat/75}{Astrochemistry (75)}}

\submitjournal{ApJ, \today}

\section{Introduction}\label{sec:intro}

Icy mantles on interstellar and circumstellar grains are a key reservoir of volatile elements and represent the primary physical link through which solid phase material is inherited from molecular clouds by planet-forming disks \citep{Gibb_2004, Aikawa_2012, Goto_2021}. Infrared ice bands provide direct access to the solid-phase inventory of abundant species such as H$_2$O, CO$_2$, and CO, and at higher spectral fidelity, to minor constituents, ions, and complex organics \citep{McClure_2023, Dishoeck_2025}. Over the past three decades, observational progress has been driven by successive space-based facilities. ISO established a foundational inventory of major ices toward a limited set of luminous, embedded sources \citep{Gibb_2004}.
Spitzer expanded sensitivity to solar-type systems and enabled 
systematic mid-infrared surveys \citep{Boogert_2008, Pontoppidan_2008}, detecting H$_2$O and CO$_2$ ice via bending modes at 6.0--15.3~\micron\ longward of Spitzer's 5.3~\micron\ blue cutoff; solid CO was not covered.
AKARI provided critical access to the 2.5--5~$\mu$m window, capturing fundamental stretching modes inaccessible from the ground \citep{Aikawa_2012}. 

Most recently, JWST has transformed the field through its combined sensitivity and spectral resolution, enabling detailed studies of ice composition, grain-growth signatures, and disk radiative-transfer effects \citep{Sturm_2023, Dartois_2024, Smith_2025, McClure_2025}.

Despite these advances, many central questions remain difficult to address with existing samples. Ice measurements are often limited by heterogeneous target selection, modest sample sizes, and interpretation degeneracies introduced by continuum placement, mixture- or temperature-dependent band strengths, and radiative-transfer effects in structured systems (envelopes and disks). In particular, scattered-light ``band filling'' and profile warping by grain growth can make the observed optical depths of the deepest ice bands a lower limit to the true column density in disk-dominated geometries, complicating direct abundance inference without detailed modeling \citep{Sturm_2023}. Consequently, hypotheses that depend on statistical leverage, such as the cross-environment uniformity of bulk ice ratios among distinct star-forming clouds, the inheritance versus chemical reset of volatile compositions across protostellar evolutionary stages, the timing of volatile accumulation along evolutionary sequences, and the prevalence of rare chemical states, remain only weakly tested \citep{Boogert_2015, Ashby_2023}.

SPHEREx provides uniform all-sky spectroscopy from 0.75 to 5.0~$\mu$m, enabling a complementary approach to the JWST era. Its spatial and spectral resolution cannot disentangle disk substructure or identify weak bands from complex organics, but the all-sky completeness and uniformity are unprecedented for the fundamental near-infrared diagnostics of bulk ices. This paper presents a search for molecular ice signatures in 20{,}292 young stellar objects selected from the Gaia-based kinematic catalog of \citet{Kounkel_2019b}. Our goal is not to derive precise ice columns for individual sources, but to construct empirically validated detection indices and candidate catalogs that characterize ice detection rates across a kinematically-selected sample and probe correlations between ice signatures, local extinction, and observable proxies of evolutionary state, and to identify high-value targets for follow-up spectroscopy.

This paper is organized as follows. Section~\ref{sec:methodology} describes the target selection, SPHEREx spectrophotometry, the template-fitting framework, and the laboratory ice spectra adopted for our molecular species. The validation strategy, including the ice-confirmed sample, the control experiments, and detection metrics is developed in Section~\ref{sec:selection}. Section~\ref{sec:results} presents the candidate census, multi-band detections, recovery of literature-confirmed sightlines, and the demographic trends with extinction, SED morphology, and age. Section~\ref{sec:conclusions} summarizes the principal results.

\section{Methodology} \label{sec:methodology}

\subsection{Sample Selection} \label{sec:sample}

Our parent sample is drawn from the Gaia DR2-based kinematic census of \citet{Kounkel_2019b}, who applied HDBSCAN clustering to five-dimensional astrometric phase space -- sky position, parallax, and proper motions in physical velocity units, within the Galactic midplane ($|b|<30^{\circ}$) within $\sim$1 kpc ($\varpi>1$ mas). Their catalog contains $\sim$1900 groups comprising $\sim2.9\times10^{5}$ stars with group-level age estimates (precision $\sim$0.15 dex) for each structure. \citep[see][]{Kounkel_2019b}.

From this catalog we select all groups with mean ages younger than 20~Myr, yielding 20,292 YSOs dominated by the nearest star-forming complexes (Orion, Taurus, Perseus, Chamaeleon, Vela~OB2, and Upper~Sco; Table~\ref{tab:theia_groups}). We adopt the published Gaia coordinates for all members and use the \textit{Theia} identifier as the primary grouping label for region-level demographic analyses.

Because our parent catalog is Gaia-selected, it is not expected to be complete for the most heavily embedded populations. In particular, deeply extincted Class~0/I sources may be underrepresented if they are not detected by Gaia or fail the quality cuts required for reliable kinematics. Our analysis should therefore be interpreted as a homogeneous census of ice-candidate signatures among optically accessible, nearby pre-main-sequence populations, but not as a complete inventory of the deeply embedded protostellar phase.\\

\begin{table}[t]
\centering
\caption{
Mapping between \textit{Theia} group identifiers and their associated star-forming regions
from Kounkel \& Covey (2019). Ages correspond to the mean logarithmic ages reported in their catalog.
}
\label{tab:theia_groups}
\begin{threeparttable}
\begin{tabular}{ccc}
\hline
\hline
Theia ID & Associated name & $\log(\mathrm{age/yr})$ \\
\hline
1  & LDN 988e        & 6.14 \\
2  & Chamaeleon I    & 6.28 \\
3  & FSR 0261        & 6.28 \\
4  & ASCC 125        & 6.27 \\
5  & NGC 7023        & 6.37 \\
6  & ---             & 6.38 \\
7  & Taurus          & 6.50 \\
8  & ---             & 6.50 \\
9  & IC 1396         & 6.64 \\
10 & ---             & 6.80 \\
11 & LDN 1251        & 6.80 \\
12 & ---             & 6.79 \\
13 & Orion           & 6.89 \\
14 & ---             & 6.90 \\
15 & ---             & 6.85 \\
16 & ---             & 6.90 \\
17 & IC 348          & 6.93 \\
18 & ASCC 122        & 6.99 \\
19 & ---             & 6.96 \\
20 & ---             & 6.97 \\
21 & NGC 1333        & 7.06 \\
22 & Vela OB2        & 7.08 \\
23 & Serpens         & 7.08 \\
41 & NGC 2264        & 7.10 \\
45 & Upper Sco       & 7.20 \\
\hline
\end{tabular}
\begin{tablenotes}
\footnotesize
\item \textbf{Notes.} The \textit{Theia} identifiers correspond to kinematic structures reported by
Kounkel \& Covey (2019). For some groups no commonly used name is provided in the original catalog; these are indicated by dashes.
\end{tablenotes}
\end{threeparttable}
\end{table}

\subsection{SPHEREx spectrophotometry} \label{sec:spherex}

For each target, we retrieve SPHEREx 0.75--5$~\micron$ spectrophotometry from the public IRSA Spectrophotometry Tool\footnote{\url{https://irsa.ipac.caltech.edu/applications/spherex/tool-spectrophotometry}}, which performs forced PSF photometry at fixed reference positions on the Quick Release~1 (QR1) \citep{SPHEREx_QR1, Akeson_2025, Bock_2025}. The instrument delivers 102 spectral channels across six linear-variable-filter (LVF) detectors with a pixel scale $\approx6\farcs15$ and resolving power $R\simeq35$--130 (wavelength-dependent). Each sky position is observed repeatedly; the tool returns the flux densities, uncertainties, wavelengths, and quality flags of every visit, and we use all valid measurements of each source. Fluxes are provided in $F_{\nu}$ (typically $\mu$Jy) and converted to $\lambda F_{\lambda} = (c/\lambda)\,F_{\nu}$ for visualization and for continuum modeling. All data analyzed here were obtained during the first SPHEREx observing cycle (2025 May-December).

\subsection{Ice template fitting framework}
\label{sec:ice_fitting}

To minimize sensitivity to global SED curvature and coupling between absorption and broad infrared excess, all fitting is performed within restricted wavelength windows centered on each molecular band. We model the observed flux using a Beer--Lambert formalism:
\begin{equation}
F_{\rm obs}(\lambda) = F_{\rm cont}(\lambda)\,\exp\!\left[-\tau(\lambda)\right]
\end{equation}
where $\tau(\lambda)$ is the optical depth profile of the ice band and $F_{\rm cont}(\lambda)$ is the local continuum defined in Section~\ref{sec:continuum_def}. Each laboratory template from the LIDA database\footnote{\url{https://icedb.strw.leidenuniv.nl/spectrum_data}} is processed and normalized to unit peak optical depth as described in Section~\ref{sec:lab_templates}, such that:
\begin{equation}
\label{eq:tau_template}
\tau(\lambda) = \tau_{\rm peak}\,T(\lambda)
\end{equation}
where $T(\lambda)$ is the normalized laboratory template (\S\ref{sec:lab_templates}) and $\tau_{\rm peak}$ is the fitted amplitude recovered by the pipeline.

\begin{figure*}[t]
    \centering
    \includegraphics[width=\textwidth]{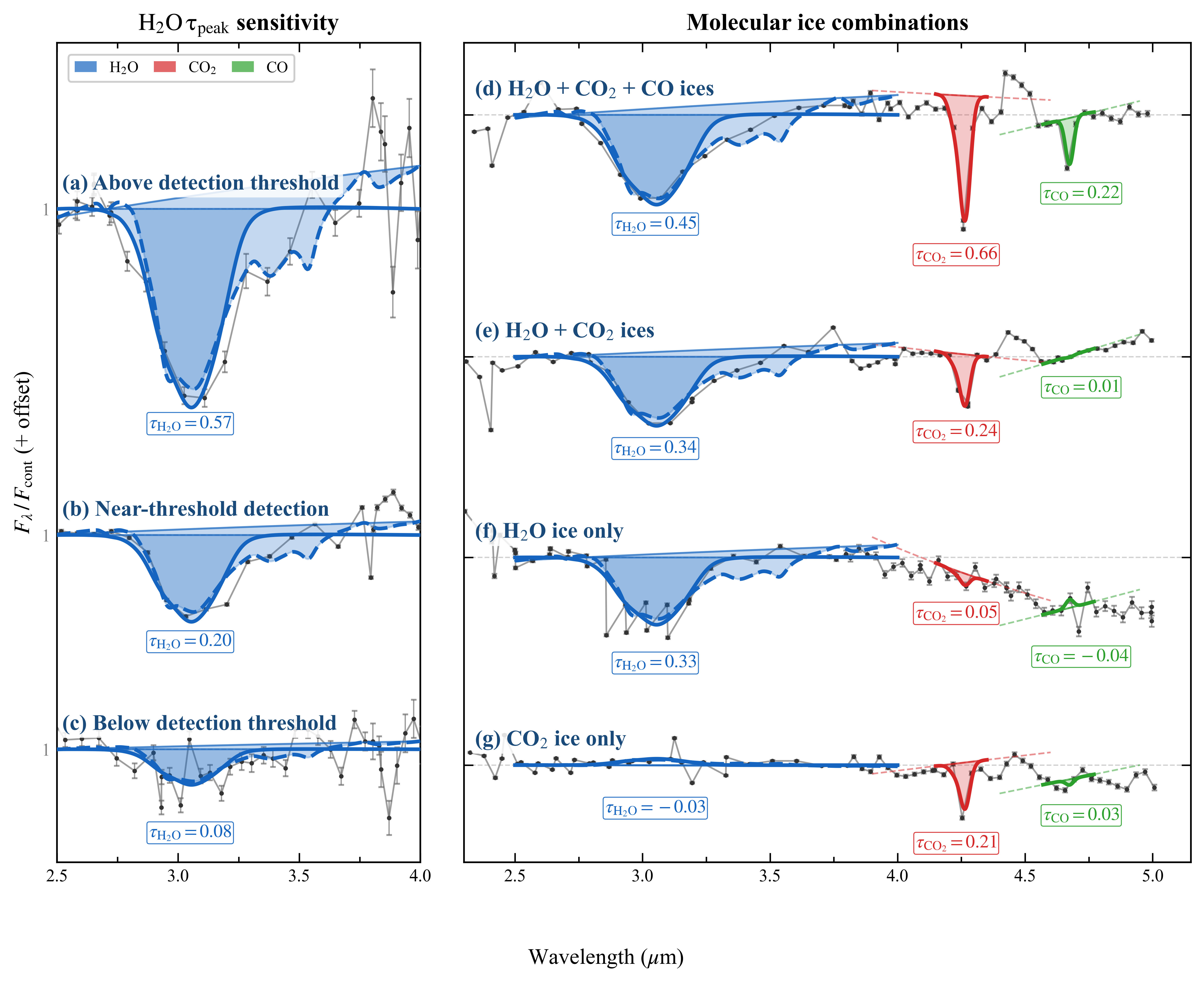}
    \caption{Representative SPHEREx spectra of molecular ice detections. In all panels, black points with grey error bars  are SPHEREx data, shown as $F_\lambda/F_{\rm cont}$ between 2.5 and 5~\micron. Each spectrum is normalized to the power-law continuum of the pure-ice H$_2$O fit, extrapolated to all wavelengths, and offset vertically; the horizontal gray dashed line marks $F_\lambda/F_{\rm cont}=1$ for each spectrum. For H$_2$O, a solid blue line shows the best-fit pure-ice template (H$_2$O at 45\,K), and a dashed blue line shows an H$_2$O-rich mixture template (H$_2$O$+$NH$_3$$+$CH$_3$OH$+$CO$_2$ at 51\,K); the reported $\tau_{\rm peak}$ corresponds to whichever template yields the lower reduced $\chi^2$. Dashed red and green lines show the local continuum fit at the CO$_2$ and CO bands, respectively. Colored fills indicate the absorption depth of each ice feature, with $\tau_{\rm peak}$ annotated at each band center. Left: Three sources spanning the $\tau_{\rm peak}$ detection range: (a)~clear detection, (b)~near-threshold, (c)~non-detection. Right: Four detection classes ordered by chemical complexity: (d)~triple-band (H$_2$O$+$CO$_2$$+$CO), (e)~H$_2$O$+$CO$_2$, (f)~H$_2$O only, and (g)~CO$_2$ only.}
    \label{fig:water_co2_co}
\end{figure*}

Figure~\ref{fig:water_co2_co} shows representative SPHEREx spectra from our sample, illustrating the H$_2$O, CO$_2$, and CO band windows, the local continuum fits, and the range of absorption depths discussed below.

\subsubsection{Local continuum definition}
\label{sec:continuum_def}
Within each band window, the SED continuum is modeled locally as a power law in wavelength,
\begin{equation}
F_{\rm cont}(\lambda) = F_0\left(\frac{\lambda}{\lambda_0}\right)^{c}
\label{eq:continuum}
\end{equation}
where $\lambda_0$ is the central wavelength of the fitting window, $F_0 = e^b$ is the continuum normalization at $\lambda_0$, and $c$ is the local spectral index. In log-flux space this becomes
\begin{equation}
\ln F_{\rm cont}(\lambda) = b + c\,\ln\!\left(\frac{\lambda}{\lambda_0}\right)
\label{eq:continuum_log}
\end{equation}
Between 2 and 5~\micron, YSO continua are well described locally by a power law in $\lambda F_\lambda$, arising from the Rayleigh--Jeans tail of the stellar photosphere combined with near infrared dust emission and scattering. This motivates a linear fit in log--log space as the continuum baseline, which captures the local spectral slope without overfitting the broad curvature characteristic of young stellar spectra.

\noindent The full model in log-flux space is therefore:
\begin{equation}
\ln F(\lambda) =
b + c\,\ln\!\left(\frac{\lambda}{\lambda_0}\right)
- \tau_{\rm peak}\,T(\lambda)
\label{eq:fullmodel}
\end{equation}
The three parameters $(\tau_{\rm peak},\,b,\,c)$ are estimated simultaneously via weighted least squares in log-flux space. For each SPHEREx channel~$i$, we define the log-flux response variable
$z_i \equiv \ln F_i$,
where $F_i$ is the observed flux.
By standard error propagation, the uncertainty in $z_i$ is
$\sigma_{z,i} = \sigma_{F,i}/F_i$,
giving per-point weights $w_i = \sigma_{z,i}^{-2} = (F_i/\sigma_{F,i})^2$.
The model is linear in $(\tau_{\rm peak},\,b,\,c)$ and is therefore
solved directly without iteration.

\subsubsection{Continuum noise and \texorpdfstring{$\tau$}{tau}-based metrics}
\label{sec:tau_metrics}

We estimate the effective noise in optical depth directly from continuum sidebands flanking each molecular band. For the \hbox{H$_2$O}\ 3.05~\micron\ band, we define reference intervals of $2.50$--$2.75$~\micron\ and $3.70$--$4.00$~\micron (blue and red sidebands, respectively). These regions were empirically selected as minimally affected by strong molecular absorption or instrumental artifacts in SPHEREx data. A single power-law continuum is fit jointly to both sideband windows, and the optical depth residuals are computed over the combined pool of channels from both windows:
\begin{equation}
\tau_{\rm cont}(\lambda) = -\ln\!\left(\frac{F_{\rm obs}}{F_{\rm cont}}\right)
\end{equation}
The noise level is estimated as the normalized median absolute deviation:
\begin{equation}
\sigma_{\tau,{\rm cont}} = 1.4826\times {\rm MAD}(\tau_{\rm cont})
\label{eq:sigmatau}
\end{equation}
which provides a robust, outlier-insensitive estimator of the per-channel optical depth noise. $\tau_{\rm cont}$ denotes the residuals pooled from both sideband windows. A floor of $\sigma_{\tau,\rm cont} \geq 10^{-3}$ is applied to prevent artificially large SNR values in nearly noiseless synthetic spectra.

\noindent The peak optical depth signal-to-noise ratio is then:
\begin{equation}
{\rm SNR}_{\tau_{\rm peak}} = \frac{\tau_{\rm peak}}{\sigma_{\tau,{\rm cont}}}
\label{eq:snr_tau}
\end{equation}

\subsubsection{Band area metric}

To complement the peak-based detection, we compute the integrated optical depth over the band in wavenumber space, following the convention used in the ice literature \citep[e.g.,][]{Boogert_2015}:
\begin{equation}
\mathcal{A} = \int_{\nu_1}^{\nu_2} \tau(\nu)\,d\nu
\label{eq:area}
\end{equation}
where $\nu = 10^4/\lambda\;[\micron]$ is the wavenumber in cm$^{-1}$, and the integration limits $(\nu_1, \nu_2)$ correspond to the band fitting window converted to wavenumber space. The optical depth profile $\tau(\nu)$ is computed from the continuum-normalized flux in the fit window using the power-law continuum derived from the sidebands. Because genuine ice absorption produces a coherent multi-channel depression, $\mathcal{A}$ provides a morphology-sensitive complement to $\tau_{\rm peak}$. Spurious single-channel excursions yield small integrated area and are naturally downweighted.

\noindent The noise on $\mathcal{A}$ is estimated under the assumption of uniform per-channel noise $\sigma_{\tau,\rm cont}$:
\begin{equation}
\sigma(\mathcal{A}) \approx \sigma_{\tau,{\rm cont}}\,\Delta\nu
\end{equation}
where $\Delta\nu$ is the wavenumber span of the integration window. The corresponding signal-to-noise ratio is:
\begin{equation}
{\rm SNR}_{\mathcal{A}} = \frac{\mathcal{A}}{\sigma_{\tau,{\rm cont}}\,\Delta\nu}
\label{eq:snr_area}
\end{equation}
${\rm SNR}_{\mathcal{A}}$ is equivalent to the ratio of the mean optical depth across the band to the per-channel noise, and therefore does not reward isolated, deep single-channel features.

\subsubsection{Equivalent width}

We additionally compute the equivalent width of each band in wavelength space:
\begin{equation}
{\rm EW} = \int_{\lambda_1}^{\lambda_2} \left[1 - e^{-\tau(\lambda)}\right] d\lambda
\label{eq:ew}
\end{equation}
with units of \micron. This is the standard spectroscopic equivalent width expressed in terms of optical depth, and reduces to $\int\tau\,d\lambda$ in the optically thin limit. In the optically thick regime ($\tau \gg 1$), EW is bounded above by the integration bandwidth $\Delta\lambda$, while $\tau_{\rm peak}$ and $\mathcal{A}$ remain formally unbounded; however, all three metrics become poorly constrained in practice when the observed flux approaches zero and the fit amplitude loses sensitivity to the true optical depth. The primary practical distinction between EW and $\tau_{\rm peak}$ is that EW integrates over the full band profile indiscriminately, whereas $\tau_{\rm peak}$ is determined via template fitting and is therefore more selective to the spectral morphology of laboratory ice absorption profiles, making it less sensitive to spectrally diffuse contributions such as photospheric water vapor in cool stars (Section~\ref{sec:phoenix}).

\noindent We compute two uncertainty estimates for EW. The first is an analytic propagation assuming uniform per-channel noise $\sigma_{\tau,\rm cont}$:
\begin{equation}
\sigma_{\rm EW}^{2} = \sum_i \left[w_i\,e^{-\tau_i}\,\sigma_{\tau,{\rm cont}}\right]^2
\label{eq:ew_err_analytic}
\end{equation}
where $w_i$ are the trapezoidal integration weights in wavelength. This estimate accounts only for photon noise and underestimates the total uncertainty when the continuum placement is uncertain. To capture continuum-placement uncertainty as well, we compute a bootstrap estimate $\sigma_{\rm EW}^{\rm boot}$. In each of $N=100$ realizations, continuum parameters $(b_s,\,c_s)$ are drawn from the weighted-least-squares covariance matrix ${\rm Cov}(b,c) = (\mathbf{X}^\top \mathbf{W}\mathbf{X})^{-1}$ and independent Gaussian noise ($\sigma_{\tau,\rm cont}$ per channel) is added to the optical-depth profile. The standard deviation of the resulting EW distribution defines $\sigma_{\rm EW}^{\rm boot}$, our adopted uncertainty estimate.

\subsection{Laboratory ice templates}\label{sec:lab_templates}

Laboratory ice spectra provide the empirical foundation for interpreting solid-state absorption features in astrophysical environments. Controlled cryogenic experiments have demonstrated that the band profile, peak position, and relative strength of vibrational modes depend sensitively on temperature, mixture composition, and ice structure (amorphous vs. crystalline) \citep{Rocha_2022}. At SPHEREx resolution, our goal is not to resolve band substructure but to capture the gross absorption morphology of major ice species.\\

For the 3.05~\micron\ O--H stretching mode of H$_2$O, we adopt multiple laboratory templates spanning different physical conditions. These include pure amorphous H$_2$O ice at low temperature \citep{Oberg_2007} and mixed-ice compositions representative of dense-cloud chemistry \citep[e.g., H$_2$O+NH$_3$+CH$_3$OH+CO$_2$ mixtures;][]{Ehrenfreund_1999}. Mixed-ice templates are particularly important because the 3~\micron\ band is often asymmetric and may exhibit a red shoulder or extended wing depending on hydrogen bonding and molecular environment \citep{Boogert_2015}. By including both pure and mixed laboratory spectra, we allow for modest profile diversity while maintaining a controlled template set.

Laboratory absorbance spectra are converted to optical depth via $\tau = \ln(10)\,A$ and normalized to unit peak optical depth, so that the fitted amplitude $\tau_{\rm peak}$ (Equation~\ref{eq:tau_template}) directly corresponds to the maximum optical depth of the band, rather than an abstract template coefficient. This choice enables a physically interpretable detection metric that can be compared across sources.\\

For CO$_2$ (4.27~\micron) and CO (4.67~\micron), we adopt laboratory spectra measured in pure and mixed phases, following experimental datasets such as those presented by \citet{VanBroekhuizen_2006} and summarized in \citet{Boogert_2015}. These bands are intrinsically narrower than the H$_2$O 3~\micron\ feature and are therefore more challenging at SPHEREx resolution. Their detection nevertheless provides an independent solid-state tracer and a critical degeneracy breaker against photospheric H$_2$O vapor (Section~\ref{sec:phoenix}).\\

All laboratory templates are interpolated onto the SPHEREx wavelength grid within each fitting window. To prevent continuum mismatches from biasing the fit, we remove a local linear baseline from each template using the edges of the fitting interval and enforce zero optical depth outside the band core. This ensures that the fitted $\tau_{\rm peak}$ parameter reflects absorption morphology rather than long-wavelength continuum curvature.

Because SPHEREx channel sampling is sparse relative to laboratory resolution, and the instrumental line-spread function is not yet publicly characterized, we evaluate templates directly on the observed wavelength grid without convolution. Tests with modest smoothing show negligible impact on high-significance detections.\\

At SPHEREx resolution, the fitted templates function as morphology detectors rather than mixture models: degeneracies between ice composition, grain shape, and radiative-transfer geometry cannot be resolved source by source. Our objective is to identify statistically significant absorption consistent with laboratory ice profiles, not to derive mixture fractions or column densities. The simultaneous detection of multiple species (H$_2$O, CO$_2$, CO) strongly favors a solid-state origin, providing the principal degeneracy breaker against photospheric contamination \citep{Boogert_2015}.

\subsection{Confirmed ice sightlines and metrics validation}\label{sec:selection}

Reliable ice detection at SPHEREx resolution requires distinguishing coherent, multi-channel absorption from isolated low-flux channels, structured noise, imperfect continuum placement, and critically photospheric water-vapor absorption in cool stars. We address these challenges through a calibrated selection strategy that combines an external ice-confirmed sample, two independent null experiments, and a systematic metric comparison.

\begin{deluxetable*}{lcccc}
\tablecaption{Literature compilation of ice-bearing sources used as a validation sample for SPHEREx ice detection.\label{tab:ice_validation}}
\tablehead{
\colhead{Source} &
\colhead{RA (deg) J2000} &
\colhead{DEC (deg) J2000} &
\colhead{Instrument/Telescope} &
\colhead{Reference/JWST Program}
}
\startdata
W3 IRS 5 & 36.419375 & 62.097791 & Spitzer, ISO & \citet{Gibb_2004, Boogert_2008}\\
Per-emb 25 & 51.656167 & 30.257735 & JWST, Spitzer & \citet{Boogert_2008, Kim_2025}, GO 1960\\
AFGL 490 & 51.911250 & 58.783639 & ISO & \citet{Gibb_2004}\\
ASR 41 & 52.213712 & 31.294386 & AKARI & \citet{Aikawa_2012}\\
Elias 1 & 64.669583 & 28.321111 & ISO & \citet{Gibb_2004}\\
Tau042021 & 65.089345 & 28.230326 & JWST & \citet{Dartois_2025}\\
DG Tau B & 66.760583 & 26.091918 & JWST, Spitzer & \citet{Boogert_2008, Delabrosse_2024}\\
HK Tau & 67.962083 & 24.404722 & AKARI & \citet{Aikawa_2012}\\
IRAS 04302+2247 & 68.318542 & 22.889083 & AKARI & \citet{Aikawa_2012}\\
HV Tau & 69.647000 & 26.177744 & AKARI & \citet{Aikawa_2012}\\
\enddata
\tablecomments{Table~\ref{tab:ice_validation} is published in its entirety in machine-readable format. A portion is shown here for guidance regarding its form and content.}
\end{deluxetable*}

\subsubsection{Evaluable spectra}
We define a spectrum as \emph{evaluable} for a given molecular band if the SPHEREx sampling within both the band window and the adjacent continuum sidebands is sufficient to (i) constrain the template fit and (ii) robustly estimate the continuum-referenced optical-depth noise $\sigma_{\tau,\mathrm{cont}}$.

For the H$_2$O 3.05~\micron\ band, a spectrum is evaluable if:

\begin{itemize}
\item $n_{\rm fit} \ge 22$ valid channels within the fitting window (2.6–3.8~\micron),
\item $n_{\rm cont} \ge 8$ valid channels across the defined sidebands (2.5–2.75 and 3.7–4.0~\micron),
\item $\sigma_{\tau,\mathrm{cont}} \le 0.2$.
\end{itemize}

For the CO$_2$ 4.27~\micron\ and CO 4.67~\micron\ bands, a spectrum is evaluable if:
\begin{itemize}
    \item $n_{\rm fit} \ge 4$ valid channels within the fitting window,
    \item $n_{\rm cont} \ge 3$ valid channels across the defined sidebands,
    \item $\sigma_{\tau,\mathrm{cont}} \le 0.2$.
\end{itemize}

These empirical thresholds effectively reject spectra dominated by structured noise or insufficient wavelength coverage for the first cycle of SPHEREx observations, as confirmed by control-sample experiments.

Throughout the rest of the paper, we report both the \emph{raw detection fraction} (detected / total sample) and the \emph{conditioned detection fraction} (detected / evaluable), where the latter quantifies the performance of each detection statistic in the regime where it is well defined.

\subsubsection{Calibration and control samples}\label{sec:control_samples}

Our primary calibration set or ``Gold Sample'' consists of 97 sources with published H$_2$O detections from four missions: ISO/SWS (\citealt{Gibb_2004}), AKARI/IRC (\citealt{Aikawa_2012,Kim_2025}), Spitzer/IRS (\citealt{Boogert_2008,Pontoppidan_2008}), and JWST/NIRSpec (\citealt{Slavicinska_2024, Arulanantham_2024,Delabrosse_2024, Smith_2025,McClure_2025,Dartois_2025, Dartois_2026}), listed in Table~\ref{tab:ice_validation}. Spanning embedded protostars, edge-on disks, and background stars, these sources confirm the same vibrational mode targeted by our pipeline and collectively define the reference $\tau_{\rm peak}$ distribution at SPHEREx resolution. 
Only 1 of the 21 ISO/SWS sightlines yields an evaluable H$_2$O spectrum, these are the most luminous sources in the compilation and largely exceed the SPHEREx dynamic range, so they are removed by the quality filters of Section~\ref{sec:selection} rather than by any mission-specific cut. Spitzer/IRS sources in this sample were identified via secondary mid-infrared features (H$_2$O at 6.0 and 6.8~\micron; CO$_2$ at 15.3~\micron), not the primary stretching bands; however, the ice columns along these sightlines are large enough that the corresponding 3.05~\micron\ absorption exceeds our detection threshold in evaluable SPHEREx spectra.
The majority are luminous, deeply embedded Class~0/I protostars without Gaia~DR3 $T_{\rm eff}$, so the gold sample enters only the $T_{\rm eff}$-independent threshold calibration.\\

To measure the empirical false-positive rate (FPR) from within the same survey data and reduction pipeline, we construct a negative control sample of weak-lined T~Tauri stars (WTTS) with negligible infrared excess and low extinction ($A_V \lesssim 0.2$) from the Orion star-forming complex \citep{Hernandez_2014, Briceno_2019}. Because WTTS lack circumstellar envelopes and span a similar temperature range to the main sample, they provide a realistic null distribution for photospheric and noise-driven false positives. 

%%%%%%%%%%%%%%%%%%%%%%%%%%%%%%%%%%%%%%%%%%%%%%%%%%%%%%%%%%%%%%%%%%%%%%%%%%
The WTTS control comprises 100 H$_2$O-evaluable and 105 CO$_2$-evaluable  spanning $T_{\rm eff} = 3330$--6000~K (median 3526~K). This temperature coverage does not reach into the regime of severest photospheric water-vapor contamination ($T_{\rm eff} \lesssim 3000$~K). FPR$_{\rm WTTS} = 1\%$ should therefore be interpreted as the contamination rate for this temperature range only; for cooler stars we rely on the PHOENIX+noise experiment (Section~\ref{sec:phoenix}; $2000 \leq T_{\rm eff} \leq 10{,}000$~K, FPR$_{\rm PH,n} = 5.5\%$). Notice that WTTS sample is regionally specific to Orion and may not capture the full photospheric diversity of the survey.\\

\begin{figure}[h!]
    \centering
    \includegraphics[width=0.5\textwidth]{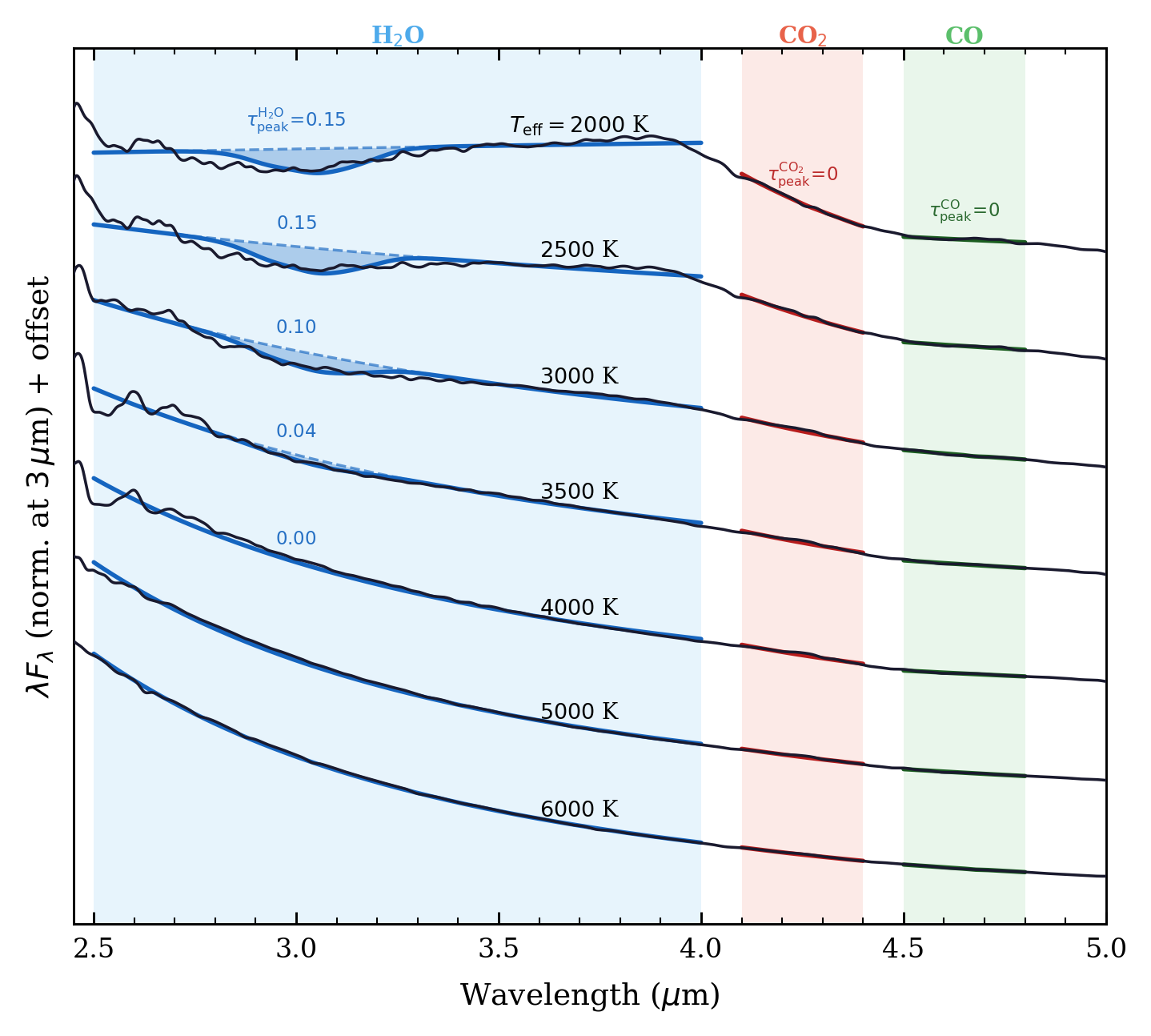}
    \caption{Photospheric water vapor absorption as a function of effective temperature. Black solid lines show the PHOENIX/BT-Settl synthetic spectra at $\log g = 4.0$, degraded to SPHEREx resolution for seven effective temperatures from 2000 to 6000~K, normalized at 3~\micron\ and offset vertically for clarity. Shaded regions mark the fitting windows, and the solid colored lines are the lab profile fit to the data. The dashed curve is the local power-law continuum fit; $\tau_{\rm peak}^{\rm H_2O}$ is the value returned by the pipeline for each model. For $T_{\rm eff} \lesssim 4000$~K, photospheric water vapor produces a measurable absorption feature within the H$_2$O window; however, all values remain below the detection threshold of $\tau_{\rm peak}^{\rm H_2O} = 0.1955$. No analogous photospheric signal appears in the CO$_2$ or CO windows at any temperature.}
    \label{fig:water_vapor}
\end{figure}

To isolate the photospheric water-vapor contribution independently of noise, we generated synthetic SPHEREx-resolution spectra from PHOENIX/BT-Settl atmospheric models \citep{Asplund_2009,Allard_2012} spanning $2000 \le T_{\rm eff} \le 10{,}000$~K in steps of 100~K, at $\log g = {3.0, 3.5, 4.0, 4.5, 5.0, 5.5}$. These models include gas-phase molecular opacity, in particular the 2.7--3.2~\micron\ water band (Figure~\ref{fig:water_vapor}), as well as CH$_4$ and TiO \citep{Barber_2006}.
Each model was degraded to the SPHEREx wavelength grid and, in a second version, augmented with correlated noise matched to the empirical $\sigma_{\tau,\mathrm{cont}}$ distribution of the survey. The noiseless set isolates photospheric signal bias; the noise-added set captures the joint effect of photospheric structure and realistic noise. The combined library contains 11,360 spectra.

\subsubsection{Metric comparison and threshold calibration}\label{sec:metrics}
For each evaluable spectrum, and independently for the H$_2$O, CO$_2$, and CO bands, we compute six candidate detection metrics: three measures of band strength such as the fitted peak optical depth $\tau_{\rm peak}$, the integrated band area $A_{\tau\nu}$, and the continuum-normalized equivalent width $EW_\lambda$, and the signal-to-noise ratio of each (${\rm SNR}_{\tau}$, ${\rm SNR}_{A}$, ${\rm SNR}_{EW}$).

Thresholds are calibrated using only two samples: the gold sample as positives and the WTTS as negatives. For a threshold $t$, the true positive rate ${\rm TPR}(t)$ is the fraction of evaluable gold-sample sources with metric $\geq t$, and $\mathrm{FPR}_{\rm WTTS}(t)$ is the fraction of evaluable WTTS spectra with metric $\geq t$. We evaluate Youden's index
$J(t) = {\rm TPR}(t) - \mathrm{FPR}_{\rm WTTS}(t)$ \citep{Youden_1950} on a fine grid of $t$ and adopt the threshold $t^{*}$ that maximizes it, $J_{\rm max} \equiv J(t^{*})$. We use the WTTS as the sole negative sample in this step because they are observed SPHEREx spectra: they carry the same instrumental systematics, continuum-placement uncertainties, and noise properties as the science sample, which synthetic spectra do not.

The PHOENIX spectra play no role in choosing $t^{*}$. Once $t^{*}$ is fixed, we apply it unchanged to the noiseless PHOENIX spectra ($\mathrm{FPR}_{\rm PH,nl}$), which isolate the purely photospheric contribution, in particular atmospheric H$_2$O absorption in cool stars and to the PHOENIX spectra with injected SPHEREx-like noise ($\mathrm{FPR}_{\rm PH,n}$), which add the effect of realistic noise. Table~\ref{tab:metrics} lists, for each metric, $t^{*}$, TPR, $\mathrm{FPR}_{\rm WTTS}$, $\mathrm{FPR}_{\rm PH,nl}$, $\mathrm{FPR}_{\rm PH,n}$, $J_{\rm max}$, and the number of H$_2$O candidates at $t^{*}$. Because the thresholds are strictly positive, solutions with $\tau_{\rm peak}<0$ are never selected as candidates. Such negative values occasionally arise when structured noise or continuum misestimation pushes the fit toward an emission-like configuration. Since thermal emission from solid H$_2$O at 3~\micron\ is not expected under typical YSO conditions, we interpret them as fitting artifacts or possible PAH emission contamination \citep{Murgia_2026, Hora_2026, Melnick_2026}.

%%%%%%%%%%%%%%%%%%%%%%%%%%%%%%%%

\begin{deluxetable*}{lccccccc}
\tabletypesize{\small}
\tablecaption{Detection metric comparison for H$_2$O (3.05~\micron) 
and CO$_2$ (4.27~\micron).\label{tab:metrics}}
\tablewidth{0pt}
\tablehead{
  \colhead{Metric} &
  \colhead{Threshold} &
  \colhead{TPR} &
  \colhead{FPR$_{\rm WTTS}$} &
  \colhead{FPR$_{\rm PH+n}$} &
  \colhead{FPR$_{\rm PH,nl}$} &
  \colhead{$N_{\rm YSO}$} &
  \colhead{$J$}
}
\startdata
\cutinhead{H$_2$O at 3.05~\micron}
$\tau_{\rm peak}$\tablenotemark{a}
  & 0.1955 & 0.965 & 0.010 & 0.055 & 0.021 &  746 & 0.955 \\
$\mathrm{Area}_{\tau\nu}$
  & 139.3 & 0.906 & 0.010 & 0.016 & 0.000 &  556 & 0.896 \\
$\mathrm{SNR}_{\tau_{\rm peak}}$
  & 5.277 & 0.969 & 0.100 & 0.078 & 0.238 & 1793 & 0.869 \\
$\mathrm{EW}_\lambda$
  & 0.085 & 0.891 & 0.030 & 0.074 & 0.079 & 2305 & 0.861 \\
$\mathrm{SNR}_{\mathcal{A}}$
  & 2.496 & 0.828 & 0.050 & 0.057 & 0.220 & 1282 & 0.778 \\
$\mathrm{SNR}_{\rm EW}^{\rm boot}$
  & 6.967 & 0.766 & 0.010 & 0.036 & 0.205 & 1562 & 0.756 \\
\cutinhead{CO$_2$ at 4.27~\micron}
$\tau_{\rm peak}^{\rm CO_2}$\tablenotemark{b}
  & 0.1418 & 1.000 & 0.000 & 0.033 & 0.000 &  724 & 1.000 \\
$\mathrm{SNR}_{\rm EW}^{\rm CO_2,boot}$
  & 2.161 & 1.000 & 0.000 & 0.007 & 0.000 &  572 & 1.000 \\
$\mathrm{EW}_\lambda^{\rm CO_2}$
  & 0.0075& 1.000 & 0.010 & 0.039 & 0.000 & 1174 & 0.990 \\
$\mathrm{Area}_{\tau\nu}^{\rm CO_2}$
  & 4.377 & 1.000 & 0.010 & 0.043 & 0.000 & 1311 & 0.990 \\
$\mathrm{SNR}_{\tau_{\rm peak}}^{\rm CO_2}$
  & 2.602 & 1.000 & 0.019 & 0.031 & 0.000 &  767 & 0.981 \\
$\mathrm{SNR}_{\mathcal{A}}^{\rm CO_2}$
  & 1.049 & 1.000 & 0.019 & 0.023 & 0.000 &  972 & 0.981 \\
\enddata
\tablecomments{All thresholds maximize $J = \mathrm{TPR} -
\mathrm{FPR}_{\rm WTTS}$. H$_2$O samples: gold $n=57$, WTTS $n=100$, PHOENIX+noise $n=10{,}235$, noiseless PHOENIX $n=341$. CO$_2$ samples: gold $n=67$, WTTS $n=105$, PHOENIX+noise $n=10{,}436$, noiseless PHOENIX $n=355$; quality filter $n_{\rm fit}^{\rm CO_2}\ge 4$. All FPR$_{\rm PH,nl}$ for
CO$_2$ are 0.000.
$N_{\rm YSO}$: detections in the 20,187-object science sample.}
\tablenotetext{a}{Adopted primary metric for H$_2$O 
(Section~\ref{sec:metrics}).}
\tablenotetext{b}{Adopted primary metric for CO$_2$.}
\end{deluxetable*}

Among all metrics evaluated, $\tau_{\rm peak}$ achieves the highest Youden index ($J = 0.955$): TPR$=96.5\%$ on the evaluable gold sample and the lowest FPR against the WTTS control ($1.0\%$). It also attains the highest area under curve (AUC) against all three null samples (Figure~\ref{fig:phoenix_teff}), confirming that it separates ice-bearing from ice-free spectra more effectively than area- or EW-based estimators at any choice of threshold. $\mathrm{Area}_{\tau\nu}$ reaches a lower false-positive rate against the PHOENIX controls at its own operating point, but recovers four fewer gold-sample sightlines; since photospheric contamination is independently suppressed by the CO$_2$ co-detection requirement (Section~\ref{sec:metrics}), we prioritize completeness and adopt $\tau_{\rm peak}$ as the primary H$_2$O metric.

Based on this metric comparison, we adopt two candidate classes stratified by photospheric contamination risk:
\begin{enumerate}
    \item \textit{High-confidence ice-mantle candidates} (class~A):
    sources with $\tau_{\rm peak}\ge 0.1955$ and simultaneous
    CO$_2$ detection ($\tau_{\rm peak}^{\rm CO_2}\ge 0.1418$) (131 sources).
    No PHOENIX model produces CO$_2$-like absorption at any $T_{\rm eff}$ (Section~\ref{sec:phoenix}). However, introducing noise on PHOENIX models similar as real data can inject a FPR$_{\rm PH+n}=0.3\%$ over all temperatures ($0.28\%$ for $3000 \le T_{\rm eff} \le 5900~K$). This class constitutes the primary science sample for all demographic analyses.
 
    \item \textit{H$_2$O-only candidates} (class~B):
    sources with $\tau_{\rm peak}\ge 0.1955$ but without CO$_2$ corroboration (615 sources). Of these, 312 have Gaia~DR3 $T_{\rm eff}$ and 311 of them lie at $T_{\rm eff} > 3000$~K. In this range, the noiseless PHOENIX test yields no false positives (0 of 279 spectra; FPR$_{\rm PH,nl}<1.1\%$ at 95\% confidence; Table~\ref{tab:teff_correction}), while the PHOENIX+noise test yields FPR$_{\rm PH+n}=4.4\%$ at $3000<T_{\rm eff}\le3500$~K, where most of these sources lie, decreasing to 1.1\% above 4000~K (Figure~\ref{fig:phoenix_teff}). For the remaining 303 sources without Gaia $T_{\rm eff}$, photospheric H$_2$O contamination cannot be bounded, the class~B FPR$_{\rm PH+n}$ spectra ranges from $\approx1\%$ above 4000~K to $24\%$ below 3000~K (Section~\ref{sec:demographics}). All 615 class~B sources are retained in the public catalog as candidates for spectroscopic follow-up, but none of them enter the demographic analysis (Section~\ref{sec:demographics}), which relies exclusively on H$_2$O+CO$_2$ co-detections.
\end{enumerate}

The $\tau_{\rm peak}$ distribution of class~B 
extends to 2.34, an order of magnitude above the maximum photospheric value produced by any PHOENIX model ($\tau_{\rm peak}^{\rm phot}=0.219$); roughly $\approx\!30\%$ of class~B sources exceed $\tau_{\rm peak}>0.25$, making a purely photospheric origin implausible for that subset.\\

We apply the same Youden-index analysis to the CO$_2$ (4.27~\micron) and CO (4.67~\micron) bands independently. For CO$_2$, $\tau_{\rm peak}^{\rm CO_2}$ achieves a perfect Youden index ($J=1.000$; Table~\ref{tab:metrics}), with TPR\,$=100\%$ and FPR$_{\rm WTTS}=0.0\%$, reflecting the absence of any photospheric CO$_2$ absorption in PHOENIX models at any effective temperature. This makes CO$_2$ an ideal solid-state degeneracy breaker for the H$_2$O detection.

For CO, the situation is fundamentally different. No CO metric achieves $J > 0.71$, and FPR$_{\rm WTTS}$ ranges from 4\% to 19\% at the Youden-optimal thresholds. This is not photospheric, noiseless PHOENIX spectra yield $\mathrm{FPR}_{\rm PH,nl}=0\%$ for $\tau_{\rm peak}^{\rm CO}$, while PHOENIX+noise reproduces the WTTS rate ($11.3\%$ vs.\ $13.6\%$), so the CO window is noise-limited at SPHEREx resolution. We therefore do not adopt any CO threshold as a primary or secondary detection criterion. Instead, we report $\tau_{\rm peak}^{\rm CO}$ as a supplementary corroborating indicator for sources already identified in H$_2$O$+$CO$_2$: among the 131 high-confidence candidates, 22 additionally show $\tau_{\rm peak}^{\rm CO}\ge0.10$ (corresponding to FPR$_{\rm PH,nl}=0.0\%$), constituting our triple-band sample.\\

A natural question is whether the H$_2$O detection threshold could be relaxed for stars with $T_{\rm eff} > 4000$~K, where photospheric bias becomes negligible (Section~\ref{sec:phoenix}; Table~\ref{tab:teff_correction}). Gaia~DR3 temperatures are available for $68.3\%$ of our sample, making such a refinement feasible in principle. 
As a lower bound, we consider $\tau_{\rm peak} \geq 0.15$, which for $T_{\rm eff} > 4000$~K lies far above the photospheric bias floor ($\tau_{\rm peak}^{\rm phot} \lesssim 0.01$; Table~\ref{tab:teff_correction}).

However, for YSOs with circumstellar disks or envelopes, photospheric veiling by accretion and disk emission systematically biases broadband $T_{\rm eff}$ estimates, 
making a temperature-stratified threshold unreliable for a significant fraction of the sample. Moreover, the incremental return is modest; applying this threshold exclusively to sources with $T_{\rm eff} > 4000$~K recovers 67 additional Class~B candidates ($+9\%$), 8 additional Class~A candidates ($+6\%$), and no additional triple-band detections. We defer this refinement to future work; catalog users with spectroscopic $T_{\rm eff}$ may apply the corrections in Appendix~\ref{sec:teff_correction} accordingly.\\

\subsubsection{Photospheric water vapor versus solid-state ice}\label{sec:phoenix}
A central systematic in this study is the wavelength overlap between the 3.05~\micron\ O--H stretching mode of solid H$_2$O and the 2.7--3.2~\micron\ ro-vibrational complex of gas-phase water in cool photospheres. At SPHEREx resolution  this photospheric feature is unresolved and produces continuum depressions that can mimic ice absorption.

\begin{figure*}[ht]
    \centering
    \includegraphics[width=\textwidth]{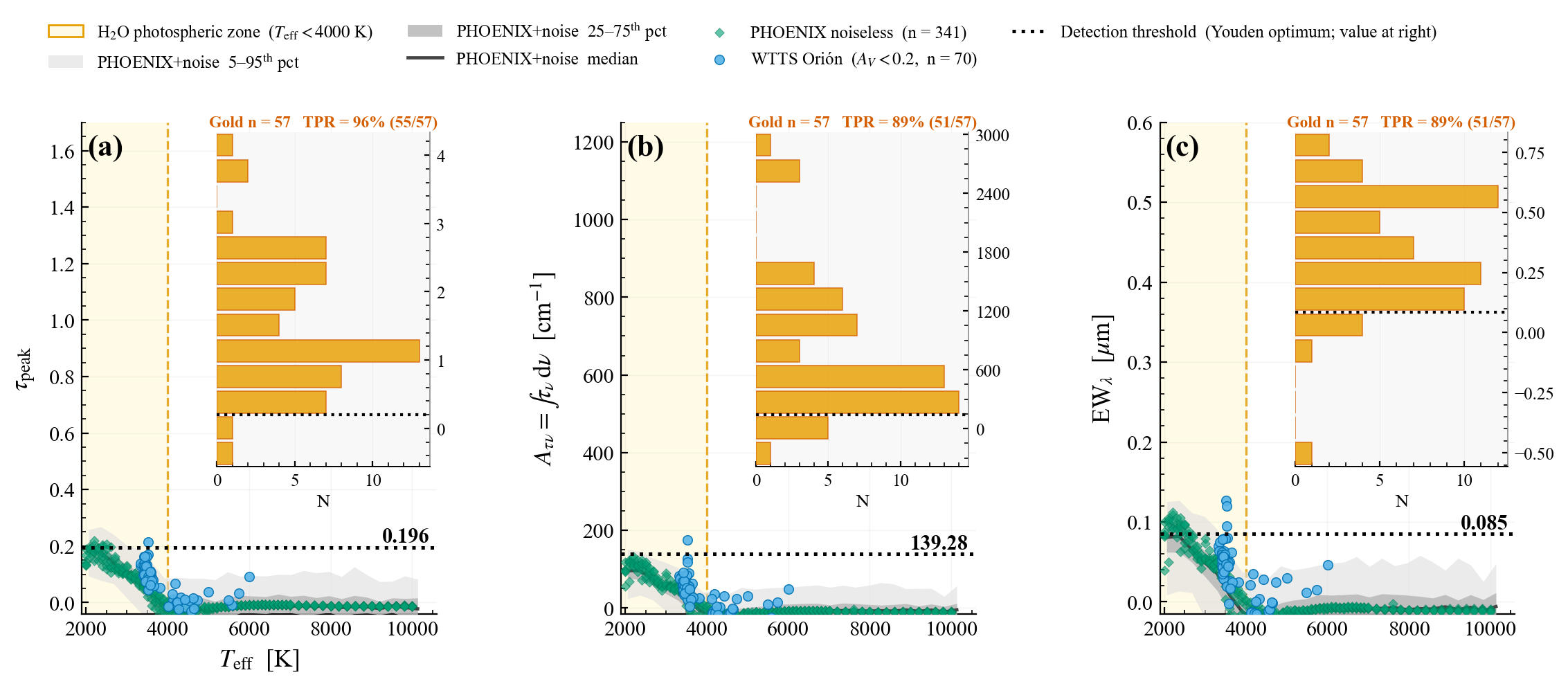}
    \includegraphics[width=\textwidth]{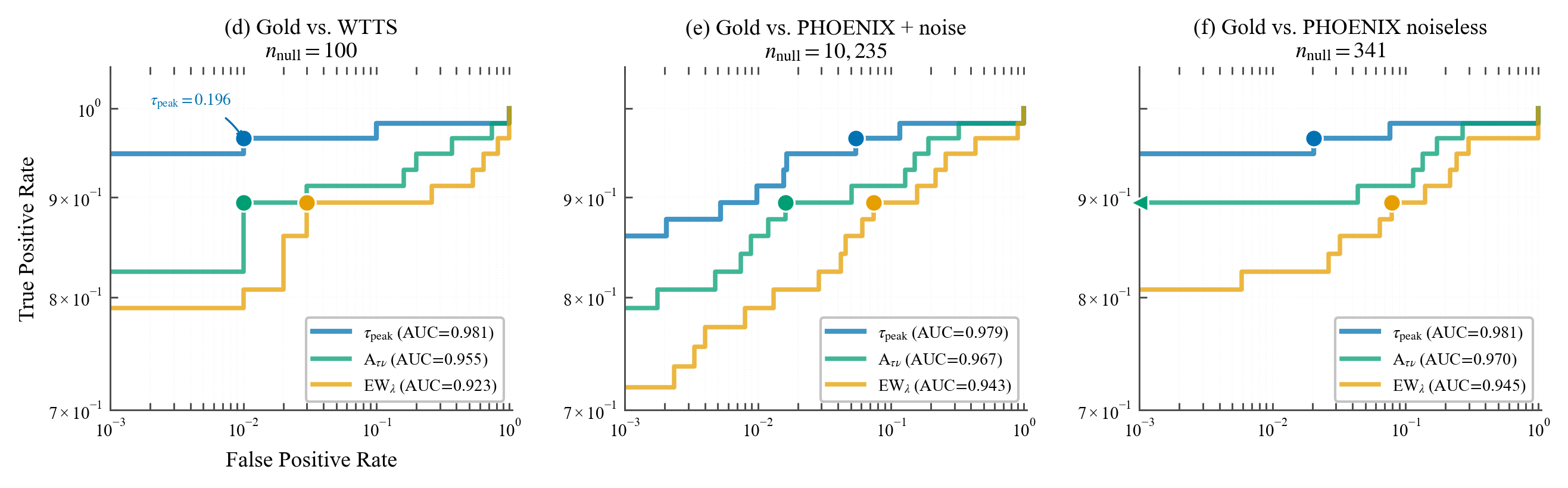}
    \caption{Photospheric contamination and metric performance for the H$_2$O band. Top row (a-c): Fitted peak optical depth $\tau_{\rm peak}$, integrated band area $A_{\tau\nu}$ and equivalent width $EW_\lambda$ as a function of $T_{\rm eff}$. The yellow region ($T_{\rm eff}<4000$~K, dashed orange line) marks the regime where photospheric water vapor is relevant. Light and dark gray bands show the 5-95$\%$ and 25-75$\%$ percentile ranges of the PHOENIX+noise spectra in 28 equal $T_{\rm eff}$ bins, and the black line is their median. Green diamonds are the noiseless PHOENIX spectra and blue circles are the Orion WTTS with Gaia~DR3 $T_{\rm eff}$ and $A_{V}<0.2$ mag. The dotted horizontal line is the optimal threshold of each metric. The orange inset histogram shows the full distribution of the metric in the evaluable gold sample, it has its own vertical axis (right), with one bin edge at the threshold, so it does not share the scale of the main panel. The dotted line in the inset repeats the threshold, and the inset title gives the true positive rate (TPR), the fraction of gold sources at or above it.
    All metrics systematically increase as temperature decreases, reflecting photospheric water-vapor absorption that mimics ice features. 
    Bottom row (d-f): Receiver operating characteristic (ROC) curves for $\tau_{\rm peak}$ (blue), $A_{\tau\nu}$ (green), and EW$_\lambda$ (orange), using the gold sample as positives and, from left to right, the Orion WTTS, the PHOENIX+noise spectra, and the noiseless PHOENIX spectra as negatives ($n_{\rm null}$ in each panel title). Axes are logarithmic to resolve the low false-positive-rate (FPR) regime, and the area under the curve (AUC) is given in each legend. Circles mark the Youden-optimal operating point of each metric (a triangle on the left axis indicates FPR$=0$, which cannot be shown on a logarithmic axis). In all three experiments $\tau_{\rm peak}$ has the highest AUC, with the ranking $\tau_{\rm peak}>A_{\tau\nu}>{\rm EW}_\lambda$.}
    \label{fig:phoenix_teff}
\end{figure*}

Figures ~\ref{fig:water_vapor}, and ~\ref{fig:phoenix_teff} illustrate this contamination.
All three metrics remain near zero above $T_{\rm eff} \gtrsim 4000$~K. Below $\approx 4000$~K, $\tau_{\rm peak}$, $EW_\lambda$ and $A_{\tau\nu}$ rise systematically with decreasing temperature; by $T_{\rm eff} \approx 3000$~K, $EW_\lambda$ reaches values of ${\approx}0.08$--$0.12~\mu$m, comparable to detections expected from moderate ice column densities. 
In contrast, $\tau_{\rm peak}$ remains well below the detection threshold ($\tau_{\rm peak}=0.1955$), across the temperature range actually spanned by our sample ($T_{\rm eff}^{\rm min}=2664$~K).

Therefore single-band H$_2$O detections in cool stars ($T_{\rm eff} \lesssim 4000$~K) require corroboration by CO$_2$ or CO before they can be attributed to ice mantles.

We note that the Spitzer/IRS ice surveys used as part of our gold sample detect H$_2$O primarily via the 6.0~\micron\ bending mode and the 12 micron libration mode \citep{Boogert_2008, Pontoppidan_2008}. Although Spitzer does not cover the 3~\micron\ stretching band directly, the ice columns along these sightlines are sufficiently large that the corresponding 3~\micron\ absorption falls well above our detection threshold when the SPHEREx spectrum is of adequate quality.

%%%%%%%%%%%%%%%%%%%%%%%%%%%%%%%%%%%

\begin{deluxetable*}{llrrr}
\tabletypesize{\footnotesize}
\tablecaption{Ice-mantle candidate classification in the main YSO sample.\label{tab:catalog_summary}}
\tablehead{
    \colhead{Class} & \colhead{Criterion} & \colhead{$N$} &
    \colhead{Fraction} & \colhead{FPR$_{\rm PH,nl}$}
}
\startdata
A & H$_2$O $\ge 0.1955$ and CO$_2$ $\ge 0.1418$ & 131 & 0.65\% & 0.0\% \\
B & H$_2$O $\ge 0.1955$, no CO$_2$ & 615 & 3.1\% & $\leq 10\%$ \\
\hline
H$_2$O (A+B) & H$_2$O $\ge 0.1955$ & 746 & 3.7\% & \\
CO$_2$ & CO$_2$ $\ge 0.1418$ & 724 & 3.6\% & 0.0\% \\
$\star$ & Class A and CO $\ge 0.10$ & 22 & 0.11\% & 0.0\% \\
\enddata
\tablecomments{Criteria are thresholds on $\tau_{\rm peak}$ of each band. Class~A (high-confidence) is the primary science sample. Class~B (H$_2$O-only) is kept in the catalog; follow-up is recommended.}
\end{deluxetable*}

%%%%%%%%%%%%%%%%%%%%%%%%%%%%%%%%%%%

\subsubsection{Visual validation (Quality Control)}

Beyond the quantitative metric analysis, all candidate sources were visually inspected in a display showing the observed spectrum, the best-fit continuum, and the ice-template fit residuals simultaneously for all three molecular bands. This way, we confirmed that high-confidence sources exhibit coherent multi-channel depressions with morphology consistent with laboratory templates, while rejected sources show asymmetric or single-channel features inconsistent with the expected band profiles.

Figure~\ref{fig:water_co2_co} shows a representative spectral atlas of selected multi-band detections from our sample, including sources with simultaneous H$_2$O and CO$_2$ absorption, and the triple-band detections (H$_2$O, CO$_2$, and CO) present concurrently.

% ==========================================================
\section{Results and Discussion}\label{sec:results}
% ==========================================================
\subsection{Incidence of H$_2$O ice candidates and catalog summary}\label{sec:incidence}

We apply the $\tau_{\rm peak}$ threshold to the science sample (20,187 objects, after removing the 105 Orion WTTS drawn from the same parent catalog and reserved as the observational null control), we identify 746 H$_2$O ice candidates, which implies a raw detection fraction of $3.7\%$ or $4.0\%$, depending on considering the full science sample or only the H$_2$O-evaluable spectra ($n=18{,}586$).
The gap between these fractions indicates that some non-detections reflect insufficient channel coverage or elevated noise rather than the physical absence of ice.

Among the H$_2$O candidates, the distribution of $\tau_{\rm peak}$ spans 0.1955--2.34, with a median value of $\tilde{\tau}_{\rm peak} = 0.226$. This range overlaps the moderate-to-strong end of the water-ice optical-depth distribution measured along embedded protostellar sightlines with JWST \citep{Smith_2025, McClure_2025}.

\begin{figure*}[ht!]
    \centering
    \includegraphics[width=\textwidth]{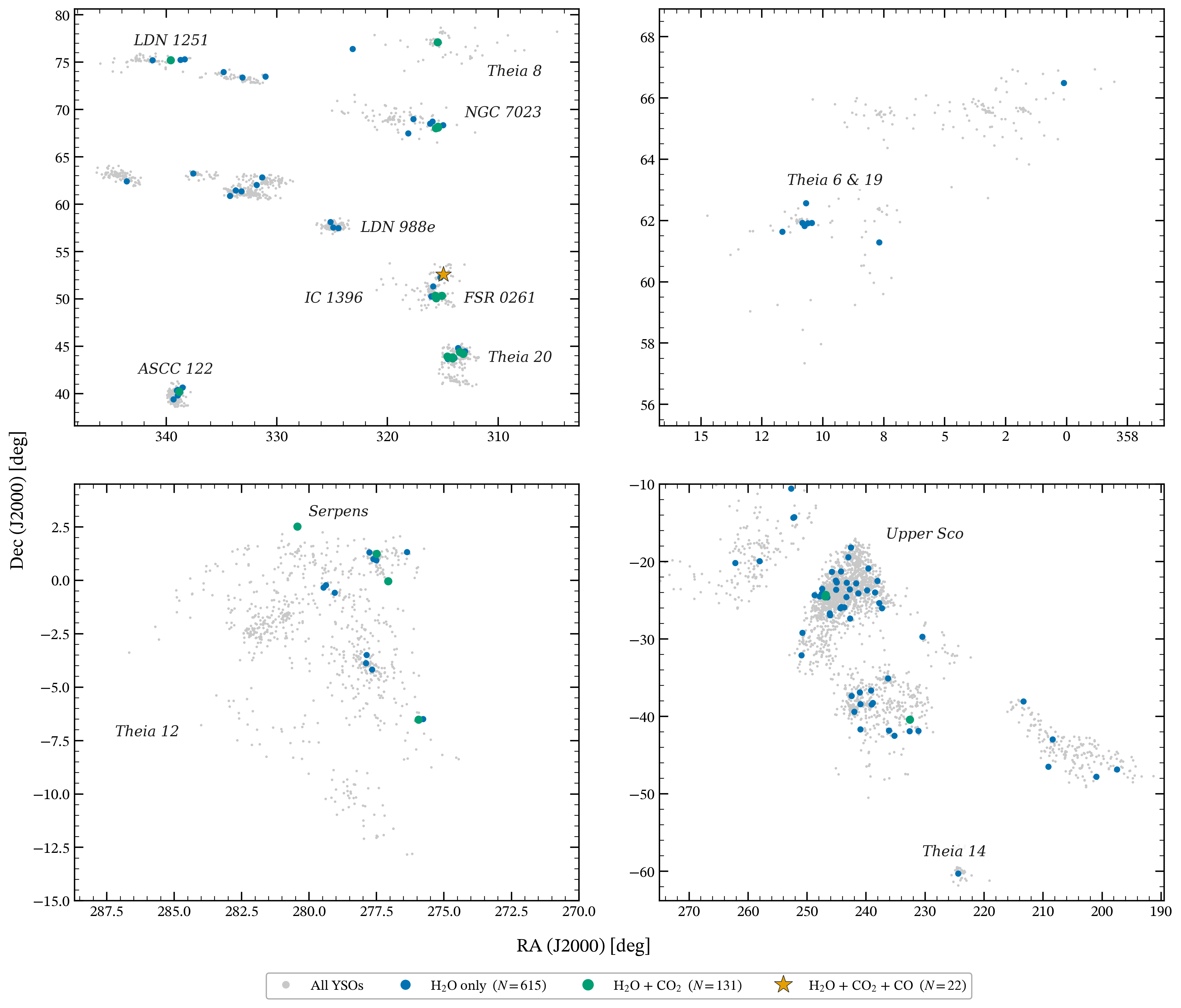}
    \caption{Sky distribution of ice-mantle candidates (Section~\ref{sec:incidence}, Table~\ref{tab:catalog_summary}) across four of the eight fields to star-forming complexes in our SPHEREx sample, drawn from the \citet{Kounkel_2019b} kinematic members catalog: the Cepheus/Cassiopeia complex (top left), Theia~6~\&~19 (top right), Serpens/Theia~12 (bottom left), and Upper~Sco/Theia~14 (bottom right). Gray points show all sources in each field; colored points show H$_2$O-only candidates (blue), H$_2$O+CO$_2$ candidates (green), and the subset of the latter with additional CO absorption (gold stars).} \label{fig:skymap1}
\end{figure*}

\begin{figure*}[ht!]
    \centering
    \includegraphics[width=\textwidth]{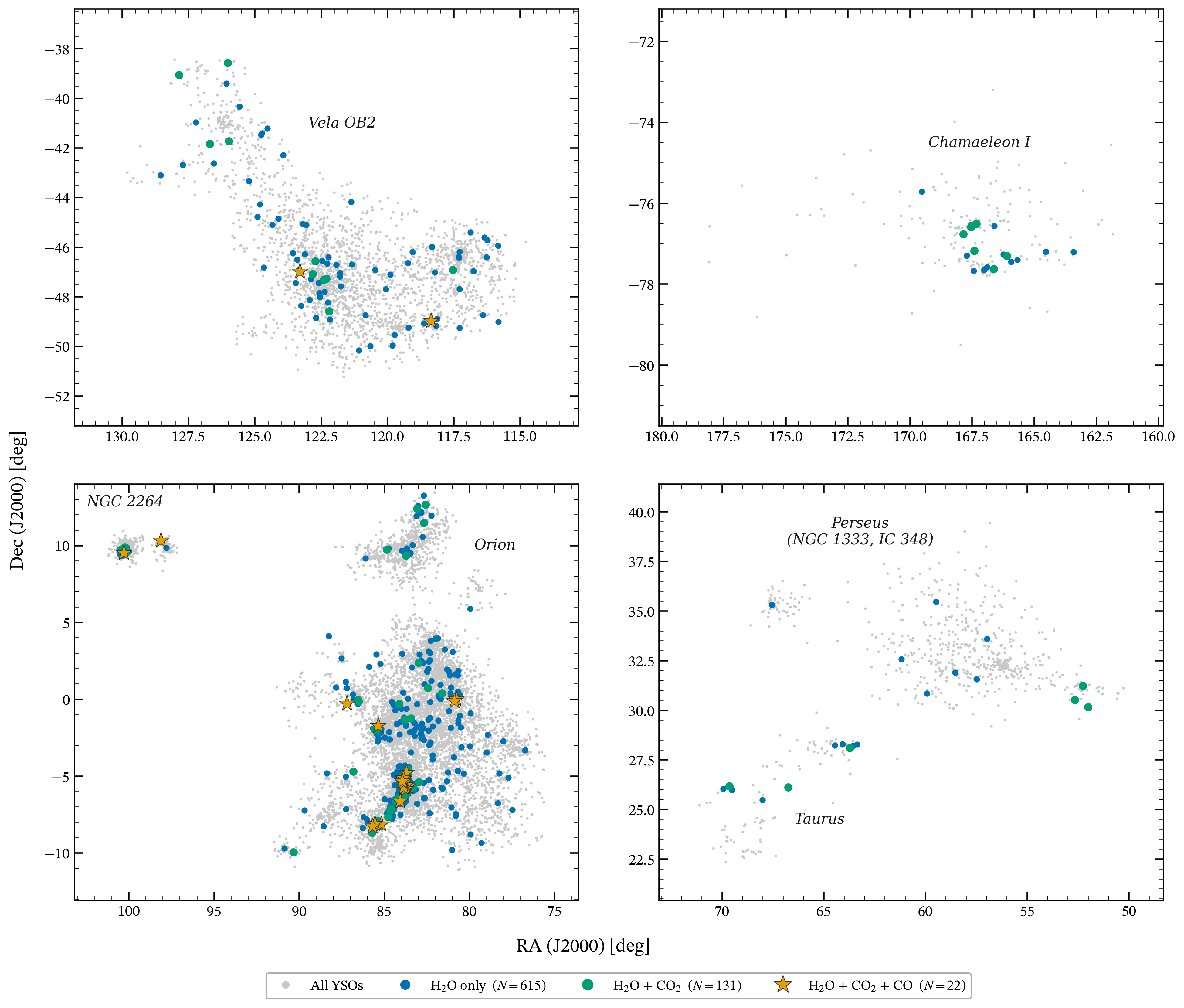} 
    \caption{Same as Figure~\ref{fig:skymap1}, but for the remaining four fields: Vela~OB2 (top left), Chamaeleon~I (top right), Orion/NGC~2264 (bottom left), and the Perseus/Taurus (bottom right).}
\label{fig:skymap2}
\end{figure*}

Table~\ref{tab:catalog_summary} summarizes the full catalog in terms of detection categories, and Figures~\ref{fig:skymap1}--\ref{fig:skymap2} show the resulting sky distribution of H$_2$O-only, H$_2$O+CO$_2$, and triple-band candidates across the star-forming complexes in our sample. We provide the catalog as a machine-readable online table with all metric columns for community use (Table \ref{tab:catalog}).

\subsection{Multi-band detections and high-confidence candidates}

The lower incidence of CO$_2$ and CO detections relative to H$_2$O is physically expected. Ice formation in dense clouds proceeds sequentially
with extinction depth \citep{Boogert_2015}. H$_2$O and CO$_2$ share a formation threshold near $A_V \sim 1.6$~mag, reflecting their close
chemical connection, CO$_2$ forms within the water-ice mantle during the earliest stages of grain processing. Our survey probes sightlines with
$A_G \leq 6.8$~mag, a regime where H$_2$O+CO$_2$ co-detections are physically expected even in the absence of CO ice.

Solid-state CO can start to freeze out beyond $A_V \gtrsim 3$~mag and undergoes catastrophic freeze-out at $A_V \gtrsim 9$~mag in the densest and coldest pre-stellar cores \citep{Boogert_2015}. The paucity of high-confidence CO signatures reflects both the noise limited CO window and the column densities accessible at $A_G \leq 6.8$~mag.

Combining all sources with H$_2$O detection plus CO$_2$ corroboration, we identify 131 high-confidence ice-mantle candidates ($0.65\%$ of the full YSO sample), retained as the primary science sample for all demographic analyses (Section~\ref{sec:demographics}). Among these, the 22 triple-band detections (H$_2$O$+$CO$_2$$+$CO; Figure~\ref{fig:water_co2_co}) are the highest-confidence ice-mantle identifications in our sample, simultaneous morphological agreement with three independent laboratory templates in distinct spectral windows is incompatible with any known photospheric artifact (Section~\ref{sec:phoenix}). We caution, however, that the CO band at 4.67~\micron\ remains noise limited (Section~\ref{sec:metrics}), and that the current analysis draws exclusively on SPHEREx Cycle~1 observations (2025 May-December). Additional sky coverage from future cycles will increase the number of independent spectral channels per source and improve the ability to distinguish solid-state from gas-phase CO; these 22 sources are therefore prime targets for both future SPHEREx cycles and JWST follow-up.

\subsection{Recovery of literature-confirmed ice sightlines}

We cross-match our SPHEREx candidate catalog against the gold ice-confirmed sample (Section~\ref{sec:control_samples}). 
Among the 57 out of 97 sources that satisfy our evaluability criteria, 55 (96\%) are recovered above the Youden-optimal $\tau_{\rm peak}$ threshold, a value representative of the gold sample's $\tau_{\rm peak}$ distribution, which is biased toward high-column embedded sources. We stress that this TPR is conditional on evaluability: the 40 gold-sample sources that fail our quality filters, primarily luminous, deeply embedded Class~0/I protostars are systematically the most ice-rich objects in the calibration set. The true TPR over the full embedded population is therefore lower than 96.5\%, and the ice-incidence fractions in Section~\ref{sec:incidence} represent lower limits on the actual ice prevalence in the most embedded phase of star formation.

\subsection{Selection biases and completeness}

Our parent sample is drawn from Gaia-based kinematic memberships \citep{Kounkel_2019b}, which emphasizes optically accessible YSOs. The most deeply embedded Class~0/I objects are systematically underrepresented, they are either too extinguished for reliable Gaia detection, or too luminous for the SPHEREx flux dynamic range. Consequently, the raw ice-incidence fractions reported here ($3.7\%$ for H$_2$O, $0.65\%$ for H$_2$O$+$CO$_2$) represent lower limits relative to the fully embedded protostellar population, where ice columns are expected to be larger and ice prevalence higher.

Catalogs of embedded YSOs constructed independently of Gaia detectability, such as the \textit{Spitzer}-based census of \citet{Megeath_2012} for Orion or the compilation of \citet{Dunham_2015} for nearby molecular clouds, provide complementary access to the deeply embedded protostellar phase that is systematically missed by our parent sample; extending the present survey to Class~0/I populations using such catalogs would place stronger constraints on the ice-incidence fractions 
reported here.

At the same time, the Gaia-kinematic selection provides a homogeneous and well-characterized framework across all nearby star-forming regions simultaneously, enabling meaningful region-to-region comparisons that are free of heterogeneous target-selection biases. In this sense, our catalog defines the ice demographics of the optically accessible pre-main-sequence population, a well-defined and physically coherent reference sample for demographic studies. 

A further consideration is source confusion. SPHEREx pixels subtend $\approx 6\farcs2$ on a side, and the point-spread function FWHM ranges from $\sim 2''$ at 0.75~\micron\ to $\sim 5''$ at 5~\micron\ \citep{Ashby_2023}. In the densest cluster cores, particularly NGC~1333, Serpens, and the Orion Nebula Cluster, multiple Gaia-resolved members can fall within a single SPHEREx beam. The Level~3 forced PSF photometry employs positional priors and a generative scene model to deblend overlapping sources \citep{Akeson_2025}, but residual confusion from close companions or extended nebulosity may still contaminate the extracted spectrum. The primary effect of unresolved blending is to add continuum flux from a companion lacking ice absorption, diluting the depth of the ice feature and biasing $\tau_{\rm peak}$ downward. Source confusion therefore works against detection, implying that our candidate fractions represent lower limits in the most crowded fields. We have not applied the $6\farcs2$ isolation criterion adopted for the SPHEREx Target List of Ice Sources \citep{Ashby_2023} because our Gaia-kinematic parent sample already excludes the most embedded, deeply crowded populations; however, individual candidates in dense cluster cores should be inspected for close companions before spectroscopic follow-up.

\begin{splitdeluxetable*}{rrlrrrrBrrrrrrcccc}
\tabletypesize{\scriptsize}
\tablecaption{SPHEREx YSO ice-band catalog (science sample).\label{tab:catalog}}
\tablehead{
\colhead{RA} & \colhead{Dec} & \colhead{Gaia ID} & \colhead{$T_{\rm eff}$} & \colhead{$\log_{10}(\rm{age})$} & \colhead{$A_G$} & \colhead{$\alpha_{2-5\,\mu{\rm m}}$} &
\colhead{$\tau_{\rm peak}$} & \colhead{S/N} & \colhead{$\tau_{\rm peak}$} & \colhead{S/N} & \colhead{$\tau_{\rm peak}$} & \colhead{S/N} & \colhead{Class} &
\colhead{Eval} & \colhead{Eval} & \colhead{Eval} \\
\colhead{(deg)} & \colhead{(deg)} & \colhead{} & \colhead{(K)} & \colhead{(log(yr))} & \colhead{(mag)} & \colhead{} &
\multicolumn{2}{c}{H$_2$O} & \multicolumn{2}{c}{CO$_2$} & \multicolumn{2}{c}{CO} & \colhead{} &
\colhead{H$_2$O} & \colhead{CO$_2$} & \colhead{CO}
}
\startdata
80.7972783 & $-1.6252496$ & 3220360641682548608 & 3563 & 6.77 & 0.95 & $-2.30$ & 0.091 & 2.4 & $-0.052$ & $-1.1$ & 0.221 & 5.8 & \nodata & T & T & T \\
83.3171355 & $-6.3098282$ & 3016778294202953088 & 4927 & 6.38 & 1.82 & $-2.44$ & 0.042 & 1.3 & $-0.077$ & $-1.7$ & 0.009 & 0.2 & \nodata & T & T & T \\
83.6789276 & $-7.0638087$ & 3016533996462888192 & 3726 & 6.53 & 0.46 & $-2.52$ & 0.071 & 2.0 & $-0.040$ & $-0.8$ & 0.121 & 1.2 & \nodata & T & T & T \\
83.6927651 & $-4.7873964$ & 3209576970937813888 & 3636 & 6.45 & 2.60 & $-1.76$ & 0.383 & 3.7 & 0.198 & 1.2 & 0.128 & 1.1 & $\star$ & T & T & T \\
83.7195199 & $-6.5932388$ & 3016942366248900736 & 3414 & 6.70 & 0.69 & $-2.15$ & 0.235 & 4.1 & $-0.085$ & $-0.7$ & 0.101 & 0.6 & B & T & T & T \\
83.8577568 & $-5.1323611$ & 3209528905962363904 & 6069 & 6.26 & 2.76 & $-0.77$ & 0.348 & 12.4 & 0.405 & 3.5 & 0.051 & 0.3 & A & T & T & T \\
118.3553116 & $-48.9845766$ & 5517412535740473344 & 3469 & 7.23 & 1.12 & $-1.97$ & 0.217 & 4.2 & 0.209 & 1.6 & 0.263 & 1.6 & $\star$ & T & T & T \\
208.3643445 & $-42.9665621$ & 6108837783037635712 & 3000 & 7.25 & 1.11 & $-2.16$ & 0.207 & 5.9 & 0.086 & 1.6 & $-0.050$ & $-2.0$ & B & T & T & T \\
315.0723173 & 50.3278905 & 2168935944493669120 & 4597 & 6.40 & 4.62 & $-1.24$ & 0.207 & 2.5 & 0.219 & 3.5 & $-0.285$ & $-6.2$ & A & T & T & T \\
\enddata
\tablecomments{A portion is shown here for guidance regarding its form and content. The full table contains all 20,187 sources of the science sample, including non-detections. Coordinates are J2000 (RA, Dec.) in degrees of the \citet{Kounkel_2019b} entry, and Gaia ID is the \textit{Gaia}~DR2 \texttt{source\_id} listed in that catalog. $T_{\rm eff}$ and $A_G$ are from \textit{Gaia}~DR3, ages from Sagitta, and $\alpha_{2-5\,\mu{\rm m}}$ is the SED slope between 2 and 5~\micron. $\tau_{\rm peak}$ and S/N are the best-fit peak optical depth and its signal-to-noise ratio in each band. Eval: T if the spectrum is evaluable in that band (Section~\ref{sec:selection}), F otherwise; classes are assigned only for evaluable bands. Class A: H$_2$O and CO$_2$ detected ($\tau_{\rm peak}({\rm H_2O})\geq0.1955$ and $\tau_{\rm peak}({\rm CO_2})\geq0.1418$); B: H$_2$O only; $\star$: satisfy class A plus CO detection simultaneously with $\tau_{\rm peak}({\rm CO})\geq0.10$; \nodata\ in the Class column: no detection among the evaluable spectra; \nodata\ in other columns: value not available. CO is a supplementary indicator and is not used to define class A or B. Negative $\tau_{\rm peak}$ reflect fitting noise around zero absorption and are never counted as detections.}
\end{splitdeluxetable*}

\begin{figure*}[ht!]
    \centering
    \includegraphics[width=0.62\textwidth]{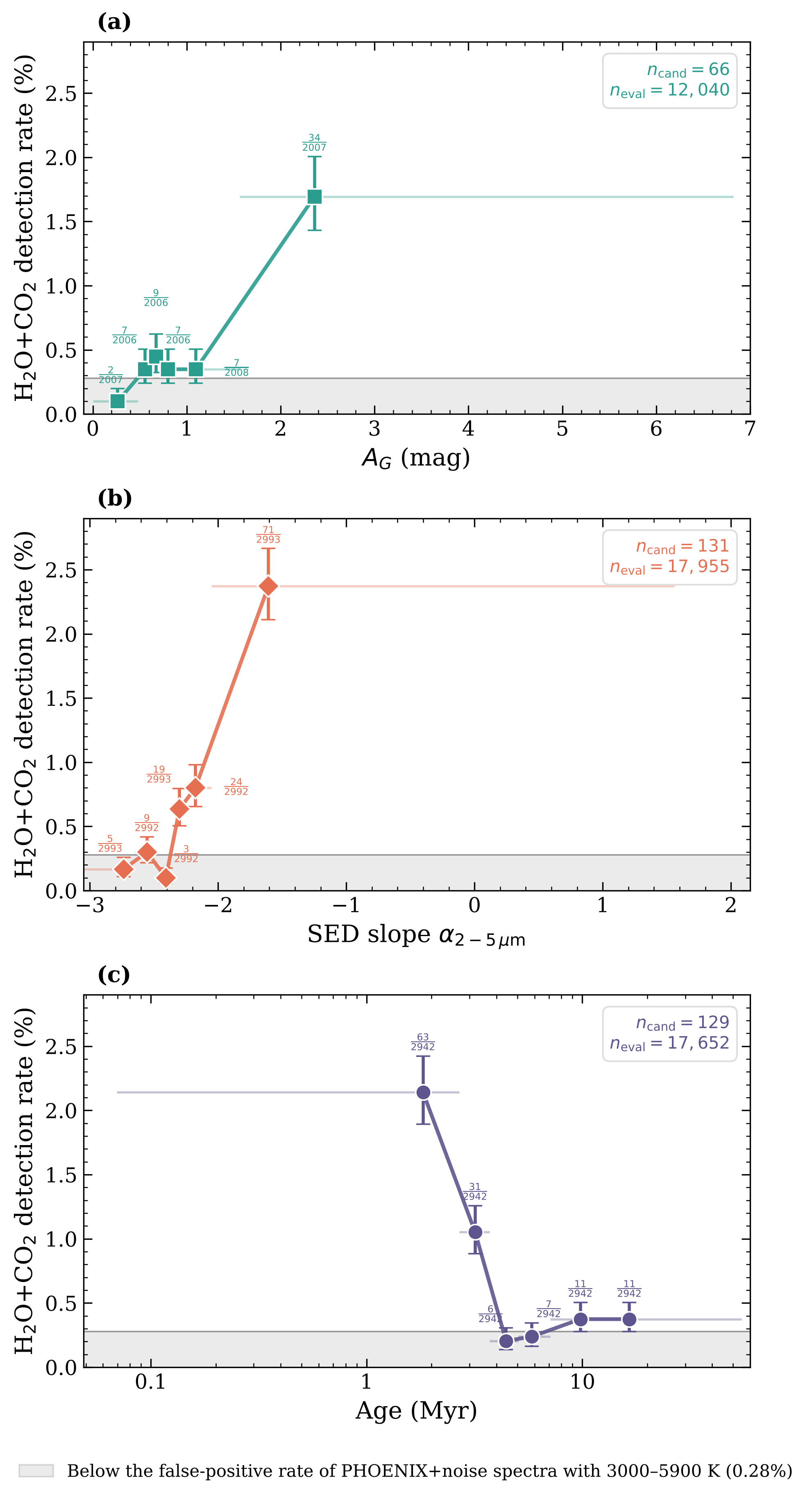}
    \caption{H$_2$O+CO$_2$ (Class~A) detection rate ($n_{\rm cand}/n_{\rm eval}$) as a function of (a) line-of-sight extinction $A_G$, (b) SED slope $\alpha_{2-5\,\mu\rm m}$, and (c) age, in six equal-population bins of the sources evaluable in both bands. Vertical error bars are 68\% Wilson confidence intervals; horizontal bars show the bin width, and the wide outer bins reflect the low source density at high extinction, steep slopes and early ages. Fractions next to each point give the number of candidates over the number of evaluable sources in the bin. The gray band extends up to the H$_2$O+CO$_2$ false-positive rate of PHOENIX+noise spectra with $3000\le T_{\rm eff}\le5900$~K ($0.28\%$; Section~\ref{sec:demographics}). All panels share the same vertical scale. Robustness against the number of bins is shown in Appendix~\ref{app:robustness}.
    }
    \label{fig:demographics}
\end{figure*}

\subsection{Demographic trends}
\label{sec:demographics}

A central motivation for anchoring this survey is the possibility of exploring evolutionary indicators across thousands of YSOs simultaneously. We characterize how the incidence of ice-mantle candidates varies with (i) line-of-sight extinction in the \textit{Gaia} $G$-band ($A_G$), (ii) the near-infrared spectral index $\alpha_{2-5\mu m}$ evaluated between 2 and 5~\micron, and (iii) individual ages.

For these analyses, we adopt the H$_2$O+CO$_2$ co-detection as the primary demographic statistic (Section~\ref{sec:metrics}), and work within the evaluable subsample of sources satisfying our quality filters for both bands simultaneously. This yields $n_{\rm eval} = 17{,}955$ sources for the SED-slope trend and $n_{\rm eval} = 17{,}652$ for the age trend, Sagitta ages being unavailable for 303 of them, while $n_{\rm eval} = 12{,}040$ sources ($67.1\%$ of the former) have a Gaia~DR3 $A_G$ measurement. Within this subsample we identify 131~H$_2$O+CO$_2$ candidates, of which 66 have an $A_G$ measurement and 129 an age estimate; the 22 triple-band detections are included in these counts. 
Detection rates are computed in six equal-population bins per variable (Figure~\ref{fig:demographics}); Appendix~\ref{app:robustness} shows that the qualitative conclusions are stable across three to ten bins.

Because Gaia $T_{\rm eff}$ is missing for 32\% of the evaluable sample and may be biased by veiling where available, we estimate the potential contamination of Class~A from the PHOENIX+noise experiment restricted to $3000\le T_{\rm eff}\le5900$~K. This interval brackets the photospheric temperatures measured from near-infrared absorption lines in Class~I and flat-spectrum protostars, which are similar to those of late-type Class~II sources and Classical T~Tauri stars \citep{Doppmann_2005, Briceno_2019}. Only 4 of the $12{,}192$ evaluable sources with Gaia $T_{\rm eff}$ are cooler than 3000~K, none of them a candidate; among the 67 class~A candidates with Gaia $T_{\rm eff}$, the coolest has 3230~K and the median is 3860~K. In this range the H$_2$O+CO$_2$ false-positive rate is $0.28\%$ (13/4665), shown as the gray band in Figure~\ref{fig:demographics}. Even below 3000~K it rises only to $0.64\%$, because no photosphere produces CO$_2$ absorption and false positives are driven mainly by noise rather than temperature.
 
We do not include any demographic trends for class~B. Without CO$_2$ corroboration, its PHOENIX+noise false-positive rate depends steeply on temperature even within $3000\le T_{\rm eff}\le5900$~K, from $4.8\%$ at 3000-3300~K to $0.95\%$ above 4000~K, and reaches $24\%$ below 3000~K. Because $T_{\rm eff}$ is unknown or veiling-biased for many sources and correlates with age, differences in the temperature mix between bins could produce or erase a trend in class~B independently of any change in the ice.
 
\subsubsection*{Extinction ($A_G$)}

The H$_2$O+CO$_2$ detection rate is $0.1$--$0.45\%$ in the five lowest-extinction bins ($A_G<1.57$~mag), at or modestly above the contamination level, and rises to $1.7\%$ (34/2007), six times that level, in the highest bin (Figure~\ref{fig:demographics}a). 
Half of the H$_2$O+CO$_2$ candidates with Gaia $A_G$ (34 of 66) reside in the highest-extinction bin. Within that bin the rise is abrupt, sources with $1.57\le A_G<2.5$~mag show $0.27\%$ (3/1127), whereas those with $A_G\ge2.5$~mag show $3.5\%$ (31/880). The canonical ice-formation threshold for H$_2$O and CO$_2$ is $A_V \approx 1.6$~mag \citep{Boogert_2015}, but the extinction at which the 3~\micron\ band becomes observable toward Taurus background stars is $A_V = 3.2 \pm 0.1$~mag \citep{Whittet_2001}, which is the quantity directly comparable to our measurement. $A_G=2.5$~mag corresponds to $A_V \approx 3$~mag for a standard extinction law with $R_V = 3.1$, in good agreement with that observed onset.

This offset is expected, the canonical threshold marks the onset of ice-mantle growth, whereas a SPHEREx co-detection additionally requires the accumulated column to exceed $\tau_{\rm peak} = 0.1955$. The extinction at which ice first forms is therefore a lower bound on the extinction at which we can detect it. The abrupt rise in co-detection rate at $A_G \gtrsim 2.5$~mag is consistent with the physical transition at which dust attenuation reduces the ambient UV field sufficiently for ice grain-surface formation to outpace photodesorption \citep{Boogert_2015, Oberg_2021}. Transport models suggest that both H$_2$O and CO are already frozen out toward the center of the cloud before the Class~0 stage begins, setting the compositional baseline that young planetary systems inherit from their natal cloud \citep{Visser_2009, Oberg_2021}.
 
\subsubsection*{SED slope ($\alpha_{2-5\micron}$)}

We define $\alpha_{2-5\,\mu\mathrm{m}} \equiv d(\log\,\lambda F_\lambda)/d(\log\,\lambda)$ evaluated between 2 and 5~$\micron$ from SPHEREx photometry, with channels inside the ice-band fitting windows masked. This index is not the traditional 2--25~$\micron$ YSO classification slope \citep{Lada_1987, Greene_1994}; because SPHEREx does not extend to mid-infrared wavelengths where warm dust emission dominates, $\alpha_{2-5\,\mu\mathrm{m}}$ is systematically more negative than the standard index for a source of equivalent evolutionary state. We use it here as an observational proxy for near-infrared circumstellar excess, more negative values indicate SEDs dominated by the stellar photosphere; values approaching zero indicate a rising near-infrared continuum from significant circumstellar material. Of the three physical variables tested, $\alpha_{2-5\,\mu\mathrm{m}}$ shows the strongest monotonic separation.

The H$_2$O+CO$_2$ detection rate is $0.1$--$0.3\%$, consistent with contamination, for $\alpha_{2-5\,\mu\mathrm{m}} \lesssim -2.35$ (Figure~\ref{fig:demographics}b), as expected for SEDs dominated by the stellar photosphere. It rises to $0.6$--$0.8\%$ at $-2.35\lesssim\alpha_{2-5\,\mu\mathrm{m}}\lesssim-2.05$ and to $2.4\%$ (71/2993) in the steepest bin, which contains 54\% of all candidates.
This concentration is consistent with the physical picture in which a significant circumstellar envelope provides the extinction, cold temperatures, and shielding from UV radiation required for ice-mantle survival \citep{Whitney_2003}.

We note, however, that $\alpha_{2-5\,\mu\mathrm{m}}$ and $A_G$ are correlated, and that our detection rates measure the prevalence of detectable ice, not the ice-to-dust ratio; a lower detection rate at small $\alpha_{2-5\,\mu\mathrm{m}}$ may simply reflect a smaller dust column, without any change in the ice content of the grains. Among the 66 class~A sources with Gaia $A_G$, $\tau_{\rm peak}$ is uncorrelated with $A_G$ (Spearman $\rho=0.08$, $p=0.50$), and the residuals of $\log\tau_{\rm peak}$ about a $\log\tau_{\rm peak}$--$\log A_G$ fit show no significant trend with age ($\rho=-0.17$, $p=0.18$). The detection threshold truncates this sample, so these tests have little power; a direct constraint on the evolution of the ice-to-dust ratio is deferred to future work.

As the envelope disperses, the near-infrared SED becomes increasingly dominated by the stellar photosphere. For a bare photosphere, $\alpha_{2-5\,\mu\mathrm{m}}$ approaches the Rayleigh-Jeans value of $-3$; applying our slope estimator to the PHOENIX/BT-Settl models over $2000\le T_{\rm eff}\le10000$~K confirms this limit. In this regime, the circumstellar material is exposed to photodesorption, which may rapidly deplete detectable ice columns \citep{Boogert_2015, Oberg_2021}. Notably, the concentration of candidates at $\alpha_{2-5\,\mu\mathrm{m}} \approx -2$ (Figure \ref{fig:demographics}b) suggests that sources retaining a residual rising near-infrared continuum represent a critical transitional regime in which a remnant envelope may still provide sufficient shielding to preserve interstellar volatiles, even as the disk becomes the dominant structural component \citep{Whitney_2003}.

\subsubsection*{Age}
The ages adopted in this section are individual per-star estimates from the Sagitta pipeline\footnote{\url{https://github.com/hutchresearch/Sagitta}} \citep{McBride_2021}, a convolutional neural network trained on Gaia and 2MASS photometry with a reported precision of $\sim$0.15~dex.

Figure~\ref{fig:demographics}c shows that the H$_2$O+CO$_2$ detection rate is $2.1\%$ (63/2942) in the youngest age bin ($\lesssim 2.7$~Myr), eight times the contamination level, falls to $1.1\%$ at 2.7--3.7~Myr, and stays at $0.2$--$0.4\%$, close to that level, at all older ages; 73\% of the candidates occupy the two youngest bins. Subdividing the sources younger than 3~Myr into three equal-population bins shows no further increase toward younger ages ($2.4\%$ below 1.7~Myr and $2.2\%$ at 1.7--2.5~Myr). The detection rate is therefore roughly constant before $\sim$2.5~Myr and declines over $\sim$2.5--5~Myr, although the $\sim$0.15~dex precision of the Sagitta ages does not allow us to locate the transition more precisely. Because the rate at older ages is comparable to or slightly above the contamination level, the data do not establish whether detectable ice disappears entirely or persists at a low level.

The decline in detection rate around 3~Myr (Figure~\ref{fig:demographics}c) is broadly consistent with the observed median lifetime of primordial protoplanetary disks \citep{Williams_2011, Hernandez_2014, Briceno_2019}, suggesting that the physical mechanism of envelope and disk dispersal could play a role. Progressive envelope dispersal reduces the dust column shielding icy grains from stellar UV radiation, accelerating photodesorption \citep{Oberg_2021}. At later evolutionary stages, vertical settling and grain growth concentrate solids toward the disk midplane, creating a strong vertical opacity gradient where residual gas and dust efficiently shield midplane ices from thermal and UV-induced desorption 
\citep{Manzo-Martinez_2020}. Radial drift of icy pebbles can further enhance this depletion by transporting icy material through volatile snowlines on timescales far shorter than viscous gas accretion, efficiently reducing the detectable outer-disk ice reservoir \citep{Williams_2025, Krijt_2025}.\\

The observed decline may also reflect active chemical processing of the solid-state volatile reservoir rather than mere dispersal. Disk midplane chemical evolution models predict that simple volatiles such as CO are converted into less volatile, more complex species, including CO$_2$ and refractory organics, on timescales of $10^5$--$10^6$~yr \citep{Eistrup_2016, Bosman_2018, Yu_2017}. Recent laboratory experiments demonstrate that UV irradiation and thermal processing applied concurrently, conditions intrinsic to early protostellar envelope evolution, promote large-radical mobility far more effectively than sequential processing pathways \citep{Sarver_2026}, implying that molecular complexity evolves on shorter timescales than previously assumed and depletes the inventory of simple ices detectable by SPHEREx as they are converted into complex organics or sequestered within grain mantles \citep{Davidsson_2021, Van_Clepper_2022, Munoz-Caro_2025}.\\

The persistence of ice signatures in a small fraction of sources with ages $>$5~Myr suggests that additional long-term pathways remain operative in a subset of systems. One candidate is delayed freeze-out in disk systems that started out warmer than average, higher stellar luminosity or disk geometry can suppress efficient ice condensation until later evolutionary stages, shifting the epoch of detectable ice formation to ages when the bulk of the population has already lost its primordial mantles. This thermal history effect is supported by gas-phase carbon depletion trends in protoplanetary disks, for instance, Herbig~Ae/Be stars at $\sim$10~Myr show outer-disk carbon depletion levels comparable to 1-2~Myr T~Tauri disks, well below the level seen in the $\sim$8~Myr T~Tauri TW~Hya, because the more luminous central objects maintain warmer disk temperatures for longer and delay CO freeze-out over a larger disk area \citep{Sturm_2022}. An analogous thermal delay could contribute to the tail of ice detections seen at $>$5~Myr; however, the Teff distribution of detected sources at these ages mirrors that of the full survey sample (median $\sim$3600~K), providing no evidence that late detections preferentially trace more massive, slower-evolving disks.\\

We report this evolutionary trend, recognizing that the observed detection rate may represent a convolved response to physical dispersal of circumstellar material and to chemical evolution of the solid-state volatile reservoir. Disentangling these two contributions is beyond the scope of the present population-scale survey, but the trend is statistically significant (Appendix~\ref{app:ag_robustness}) and motivates a dedicated follow-up with higher-resolution infrared spectroscopy or even a larger sample of YSOs.

A potential confounding factor is that younger stars tend to have steeper near-infrared SED slopes and higher extinction, both of which independently favor detection. Appendix~\ref{app:ag_robustness} repeats the analysis within three bounded bins of each control variable.
In every stratum of both SED slope and $A_G$, the youngest age bin exceeds the contamination level, by factors of 3--16 in slope and 3--27 in $A_G$. At fixed SED slope, the logistic slope of the detection probability with $\log({\rm age})$ is $-0.84\pm0.22$ per dex ($p=2\times10^{-4}$). At fixed $A_G$, available for only 66 candidates, a model linear in $\log({\rm age})$ does not detect a significant trend ($p=0.30$), whereas a model that separates ages $\lesssim3$~Myr, adopted after inspecting Figure~\ref{fig:demographics}c, gives an odds ratio of 3.1 ($p=1\times10^{-4}$), and of 2.0 ($p=0.03$) when SED slope and $A_G$ are both included as covariates. We therefore regard the age dependence as established independently of SED slope, and as consistent with, but less firmly established independently of, extinction.
 
The concordance of the H$_2$O+CO$_2$ demographics with extinction, near-infrared excess, and youth constitutes an internal self-validation; the candidates trace the physical conditions expected for ice survival.\\

The spatial distribution of H$_2$O+CO$_2$ and triple-band candidates shown in Figures~\ref{fig:skymap1}--\ref{fig:skymap2} visually corroborates these trends; candidates are not uniformly distributed across the kinematic membership of each region, consistent with the 
strong dependence of co-detection rate on line-of-sight extinction and near-infrared excess identified above. The highest-confidence sightlines serve as a prioritized target list for JWST follow-up spectroscopy.
Triple-band detections (H$_2$O$+$CO$_2$$+$CO) are concentrated in four groups: Orion ($n=17$), NGC~2264 ($n=2$), Vela~OB2 ($n=2$), and LDN~988e ($n=1$).
Per-group co-detection fractions vary across the sample but the small number of detections in most groups precludes statistically robust region-to-region comparisons; the full per-group census is available in the machine-readable catalog.\\
Together, these demographic trends establish that SPHEREx can characterize ice-mantle incidence across thousands of YSOs simultaneously, providing a statistical baseline that complements the detailed sightline chemistry accessible to JWST.

% ==========================================================
\section{Summary and Conclusions}\label{sec:conclusions}
% ==========================================================

We carried out a population-scale search for solid-state ice signatures in SPHEREx 0.75--5~\micron\ spectra of 20,292 young stellar objects ($<$20~Myr), spanning the nearest star-forming complexes. Using a Beer--Lambert laboratory-template framework, continuum-referenced optical-depth noise estimation, and a systematic metric comparison calibrated against ice-confirmed and ice-free control samples, we constructed a conservative, high-purity detection pipeline optimized for low-resolution survey data. Our principal results are the following.

\begin{enumerate}

    \item Of the six candidate metrics evaluated, the fitted peak optical depth $\tau_{\rm peak}$ maximizes both the Youden index and the area under the ROC curve against all three control samples, and is adopted as the primary detection criterion.

    \item $\tau_{\rm peak}$, Area and EW-based metrics exhibit a systematic false-positive rate that rises steeply below $T_{\rm eff} \approx 4000$~K in PHOENIX spectra, confirming that photospheric water vapor can mimic ice absorption at low spectral resolution. This is a signal bias that cannot be corrected by noise estimation. We caution the community that single-band H$_2$O detections in cool pre-main-sequence stars should be corroborated by independent solid-state tracers before being attributed to ice mantles.

    \item We identify 746 H$_2$O ice candidates ($3.7\%$ of the full YSO sample; $4.0\%$ of evaluable spectra), of which 131 are high-confidence multi-band candidates with corroborating CO$_2$ detections. Among these, 22 additionally show CO absorption, representing the highest-purity ice-mantle identifications in the sample and prime targets for JWST follow-up.

    \item Among the three physical variables explored, line-of-sight extinction $A_G$, SED slope $\alpha_{2-5\mu m}$, and age, all three show highly significant ($p \leq 10^{-8}$ in each case) departures from a uniform distribution of H$_2$O+CO$_2$ candidates. The detection rate peaks at $\sim$2.1\% for the youngest ($\lesssim$2.7~Myr) stars and declines by an order of magnitude by 5~Myr, a decline that persist at fixed SED slope, consistent with envelope dispersal and increasing photodesorption. Co-detections rise steeply above $A_G \approx 2.5$~mag, consistent with the extinction at which the 3~\micron\ ice band becomes observable in quiescent clouds. The ice-candidate population is strongly concentrated toward sources with significant near-infrared circumstellar excess ($\alpha_{2-5\,\mu\mathrm{m}} \gtrsim -2$), suggesting that envelope retention, and possibly disk structure, are the primary drivers of ice-mantle detectability at SPHEREx sensitivity.

\end{enumerate}

\section*{acknowledgments}

\noindent We acknowledge Jesús Hernández and Nuria Calvet for helpful recommendations. CRZ acknowledges support from project UNAM-DGAPA-PAPIIT IN107226. This publication makes use of data products from the Spectro-Photometer for the History of the Universe, Epoch of Reionization and Ices Explorer (SPHEREx), which is a joint project of the Jet Propulsion Laboratory and the California Institute of Technology, and is funded by the National Aeronautics and Space Administration. The authors acknowledge the use of Claude (\url{https://claude.ai}, version Sonnet 4.6 and Opus 4.6) for editing the paper.

\vspace{5mm}
\facilities{SPHEREx, and \textit{Gaia}-DR3.}

\software{Numpy \citep{Harris2020}, Matplotlib \citep{Hunter2007}, Astropy \citep{2013A&A...558A..33A}, Scipy}

\appendix

\section{Photospheric water-vapor correction for catalog users}
\label{sec:teff_correction}

We provide empirical correction functions for the photospheric water-vapor bias introduced into H$_2$O ice metrics by cool stellar atmospheres. These are derived from our PHOENIX/BT-Settl null experiment (Section~\ref{sec:phoenix}), in which noiseless synthetic spectra spanning $2000 \leq T_{\rm eff} \leq 4500$~K were processed through the identical template-fitting pipeline. The resulting temperature-dependent bias in $\tau_{\rm peak}$, $\mathcal{A}_{\tau\nu}$,  and $EW_\lambda$ is tabulated in Table~\ref{tab:teff_correction} 
(median and 16th--84th percentile per $T_{\rm eff}$ bin). A full census of ice candidates with individual photospheric corrections applied as a function of $T_{\rm eff}$ is beyond the scope of this work and is deferred to future analysis.

For computational convenience, we provide polynomial approximations to the photospheric bias over the range $2000 \leq T_{\rm eff} 
\leq 4500$~K, intended as a practical resource rather than an exact calibration. The $\tau_{\rm peak}$ bias is well described by a third-order polynomial (RMSE~$= 0.010$):
\begin{equation}
    \tau_{\rm peak}^{\rm phot}(T_{\rm eff})
        =   3.58\times10^{-11}\,T_{\rm eff}^{3}
          - 3.50\times10^{-7}\,T_{\rm eff}^{2}
          + 1.005\times10^{-3}\,T_{\rm eff}
          - 7.26\times10^{-1}
    \label{eq:teff_poly}
\end{equation}
which evaluates to $\tau_{\rm peak}^{\rm phot} \approx 0.19$ at $T_{\rm eff} = 2100$~K and to $\approx 0.05$ at 3500~K, 
becoming negligible above ${\sim}3800$~K. Analogous second-degree fits for the band-area and equivalent-width metrics are given 
in Equations~(\ref{eq:teff_poly_area})--(\ref{eq:teff_poly_ew}). Three caveats apply: (1) the corrections inherit systematic 
uncertainties in PHOENIX water-vapor opacities and should be treated as first-order estimates; (2) no correction is needed 
for CO$_2$ or CO metrics, which show no photospheric bias at any temperature; and (3) a corrected 
$\tau_{\rm peak}^{\rm corr} = \tau_{\rm peak}^{\rm obs} - 
\tau_{\rm peak}^{\rm phot}$ is physically interpretable only 
in combination with the CO$_2$ co-detection criterion for 
sources with $T_{\rm eff} \leq 4000$~K.
\begin{align}
    \mathcal{A}_{\tau\nu}^{\rm phot}(T_{\rm eff})
        &= -7.04\times10^{-7}\,T_{\rm eff}^{2}
           - 5.78\times10^{-2}\,T_{\rm eff}
           + 241.9 \label{eq:teff_poly_area}\\[4pt]
    EW_{\lambda}^{\rm phot}(T_{\rm eff})
        &= -2.93\times10^{-9}\,T_{\rm eff}^{2}
           - 3.64\times10^{-5}\,T_{\rm eff}
           + 0.191 \label{eq:teff_poly_ew}
\end{align}

\begin{deluxetable}{ccccccc}
\tablecaption{
    Photospheric water-vapor bias in ice metrics as a function of effective temperature, derived from noiseless PHOENIX/BT-Settl synthetic spectra processed through the ice-fitting pipeline.
    \label{tab:teff_correction}
}
\tablehead{
    \colhead{$T_{\rm eff}$} &
    \colhead{$N$} &
    \multicolumn{3}{c}{$\tau_{\rm peak}^{\rm phot}$} &
    \colhead{$\mathcal{A}_{\tau\nu}^{\rm phot}$} &
    \colhead{$EW_{\lambda}^{\rm phot}$} \\
    \colhead{(K)} &
    \colhead{} &
    \colhead{} &
    \colhead{} &
    \colhead{} &
    \colhead{(cm$^{-1}$)} &
    \colhead{(\micron)}\\
    \colhead{} &
    \colhead{} &
    \colhead{16th} &
    \colhead{50th} &
    \colhead{84th} &
    \colhead{50th} &
    \colhead{50th}
}
\startdata
2000 &  6 & 0.132 & 0.141 & 0.172 &  95.9 & 0.084 \\
2100 &  6 & 0.172 & 0.191 & 0.193 & 116.8 & 0.101 \\
2200 &  6 & 0.172 & 0.192 & 0.208 & 120.6 & 0.104 \\
2300 &  6 & 0.163 & 0.186 & 0.192 & 111.4 & 0.095 \\
2400 &  6 & 0.147 & 0.170 & 0.200 & 110.6 & 0.094 \\
2500 &  6 & 0.130 & 0.157 & 0.194 & 104.6 & 0.091 \\
2600 &  6 & 0.120 & 0.136 & 0.176 &  94.0 & 0.083 \\
2700 &  6 & 0.119 & 0.143 & 0.161 &  87.2 & 0.077 \\
2800 &  6 & 0.110 & 0.120 & 0.127 &  66.6 & 0.060 \\
2900 &  6 & 0.103 & 0.107 & 0.115 &  56.7 & 0.052 \\
3000 &  6 & 0.101 & 0.103 & 0.127 &  60.2 & 0.055 \\
3100 &  6 & 0.083 & 0.094 & 0.099 &  55.9 & 0.052 \\
3200 &  6 & 0.079 & 0.088 & 0.103 &  52.0 & 0.047 \\
3300 &  6 & 0.064 & 0.070 & 0.082 &  55.7 & 0.050 \\
3400 &  6 & 0.053 & 0.060 & 0.066 &  38.6 & 0.035 \\
3500 &  6 & 0.034 & 0.050 & 0.065 &  31.9 & 0.029 \\
3600 &  6 & 0.008 & 0.037 & 0.047 &  20.7 & 0.017 \\
3700 &  6 &$-0.004$& 0.012 & 0.031 &  17.5 & 0.013 \\
3800 &  6 &$-0.027$& 0.006 & 0.024 &   7.6 & 0.005 \\
3900 &  6 &$-0.033$&$-0.012$& 0.009 &  $-5.0$ &$-0.007$\\
4000 &  6 &$-0.031$&$-0.015$&$-0.005$& $-11.0$ &$-0.012$\\
4100 &  6 &$-0.040$&$-0.023$&$-0.007$& $-15.4$ &$-0.016$\\
4200 &  6 &$-0.038$&$-0.020$&$-0.010$& $-15.2$ &$-0.015$\\
4300 &  6 &$-0.040$&$-0.026$&$-0.013$& $-19.6$ &$-0.020$\\
4400 &  6 &$-0.033$&$-0.026$&$-0.018$& $-18.4$ &$-0.018$\\
4500 &  6 &$-0.032$&$-0.027$&$-0.016$& $-18.9$ &$-0.019$\\
\enddata
\tablecomments{
    Spectra processed with the standard H$_2$O evaluability filters
    ($n_{\rm fit} \geq 22$, $n_{\rm cont} \geq 8$,
    $\sigma_{\tau,{\rm cont}} \in (0, 0.2]$).
    Six PHOENIX models per $T_{\rm eff}$ bin span $\log g \in \{3.0, 3.5, 4.0, 4.5, 5.0, 5.5\}$. Polynomial fits to these values are given in Equations~(\ref{eq:teff_poly})--(\ref{eq:teff_poly_ew}),
    valid for $2000 \leq T_{\rm eff} \leq 4500$~K.
    Negative values (above $T_{\rm eff} \gtrsim 3800$~K) reflect fitting noise around zero absorption and confirm that no photospheric correction is needed in that temperature range.
    These corrections apply only to the H$_2$O 3.05~\micron\ band;
    CO$_2$ and CO bands show no measurable photospheric bias in PHOENIX
    models at any temperature.
}
\end{deluxetable}
 
\section{Robustness of Demographic Trends Against Binning Choice}
\label{app:robustness}
The demographic trends presented in Section~\ref{sec:demographics} rely on equal-population binning, which can in principle introduce artifacts when the number of bins is small (coarse resolution, masking fine structure) or large (low candidate counts per bin, amplified Poisson noise). To verify that our conclusions are not sensitive to this choice, we repeat the analysis of the H$_2$O+CO$_2$ detection rate using three, four, five, seven, eight, nine, and ten bins across each of the three physical variables; the six-bin case is shown in Fig.~\ref{fig:demographics}.

\begin{figure*}[ht!]
    \centering
    \includegraphics[width=\textwidth]{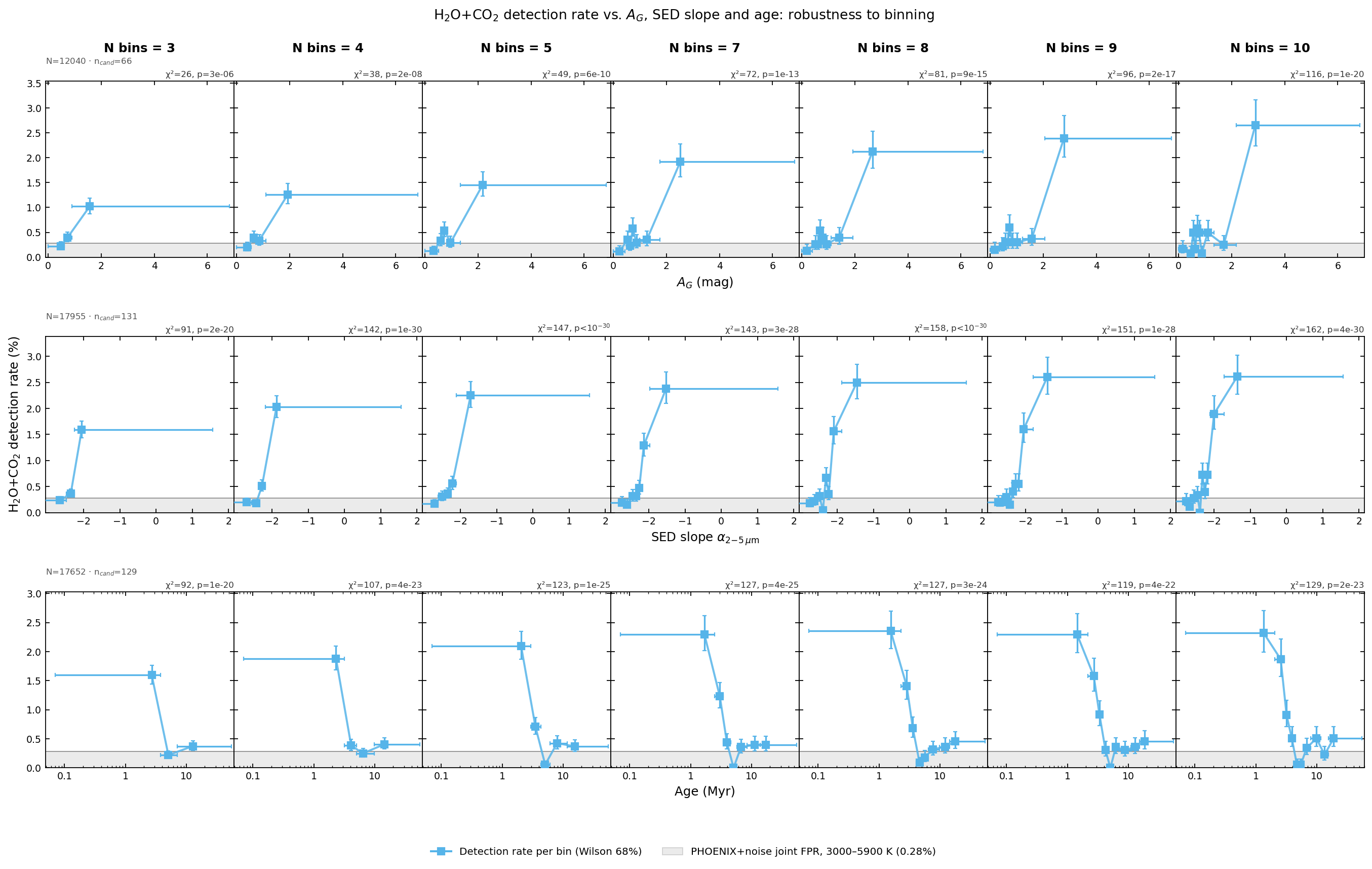}
    \caption{Robustness of the H$_2$O+CO$_2$ detection rate to the number of bins. Rows show the fraction of H$_2$O+CO$_2$ (class~A) candidates among the evaluable sources as a function of $A_G$ (top), SED slope $\alpha_{2-5\,\mu{\rm m}}$ (middle), and age (bottom). Columns repeat the analysis with 3, 4, 5, 7, 8, 9 and 10 equal-population bins (the 6-bin case is shown in Fig.~\ref{fig:demographics}). Points mark the median of each bin, horizontal bars the bin range, and vertical bars the 68\% Wilson interval. The gray band extends up to the H$_2$O+CO$_2$ false-positive rate of the PHOENIX+noise spectra (0.28\%; 3000--5900~K); rates inside it cannot be distinguished from the expected contamination. The number of sources ($N$) and candidates ($n_{\rm cand}$) with each variable available is given in the first column. Each panel lists the $\chi^2$ statistic, and the $p$-value of the test against a constant rate across bins. For every binning, the highest rates occur at high $A_G$, steep SED slope, and young age.
    }
    \label{fig:robustness}
\end{figure*}

Figure~\ref{fig:robustness} displays the results. For every combination of variable and bin count, the qualitative pattern is identical, over-representation of H$_2$O+CO$_2$ candidates at high $A_G$, steep SED slope, and young age, with corresponding under-representation in the complementary regimes. The associated $\chi^2$ statistics against a constant detection rate are highly significant in all cases ($p < 3\times10^{-6}$ at maximum, for $A_G$ with three bins; $p < 2\times10^{-20}$ for the slope and age panels at all bin counts), indicating that the detected non-uniformity does not depend on the binning scheme adopted in the main analysis.

We additionally verify that the dominant feature in each variable, the steep rise in detection rate above $A_G \approx 2.5$~mag, above $\alpha_{2-5} \approx -2.2$, and below $\sim$3~Myr is resolved with four bins and remains stable through ten bins, where the average number of candidates per bin drops to $\sim$7 for $A_G$ and $\sim$13 for the slope and age. We therefore adopt $N_{\rm bins} = 6$ as the primary binning choice for Figure~\ref{fig:demographics}, which balances statistical
stability (Wilson 68\% confidence intervals $\lesssim 0.3$ percentage points per bin) with sufficient resolution to resolve the physical transitions. The $p$-values obtained for different binnings are not independent, since they derive from the same data, and assume independent sources; they should be read as a robustness check rather than as cumulative evidence.

\section{Robustness of the Age Trend to SED Slope and Extinction}
\label{app:ag_robustness}

\begin{figure*}[ht!]
    \centering
    \includegraphics[width=\textwidth]{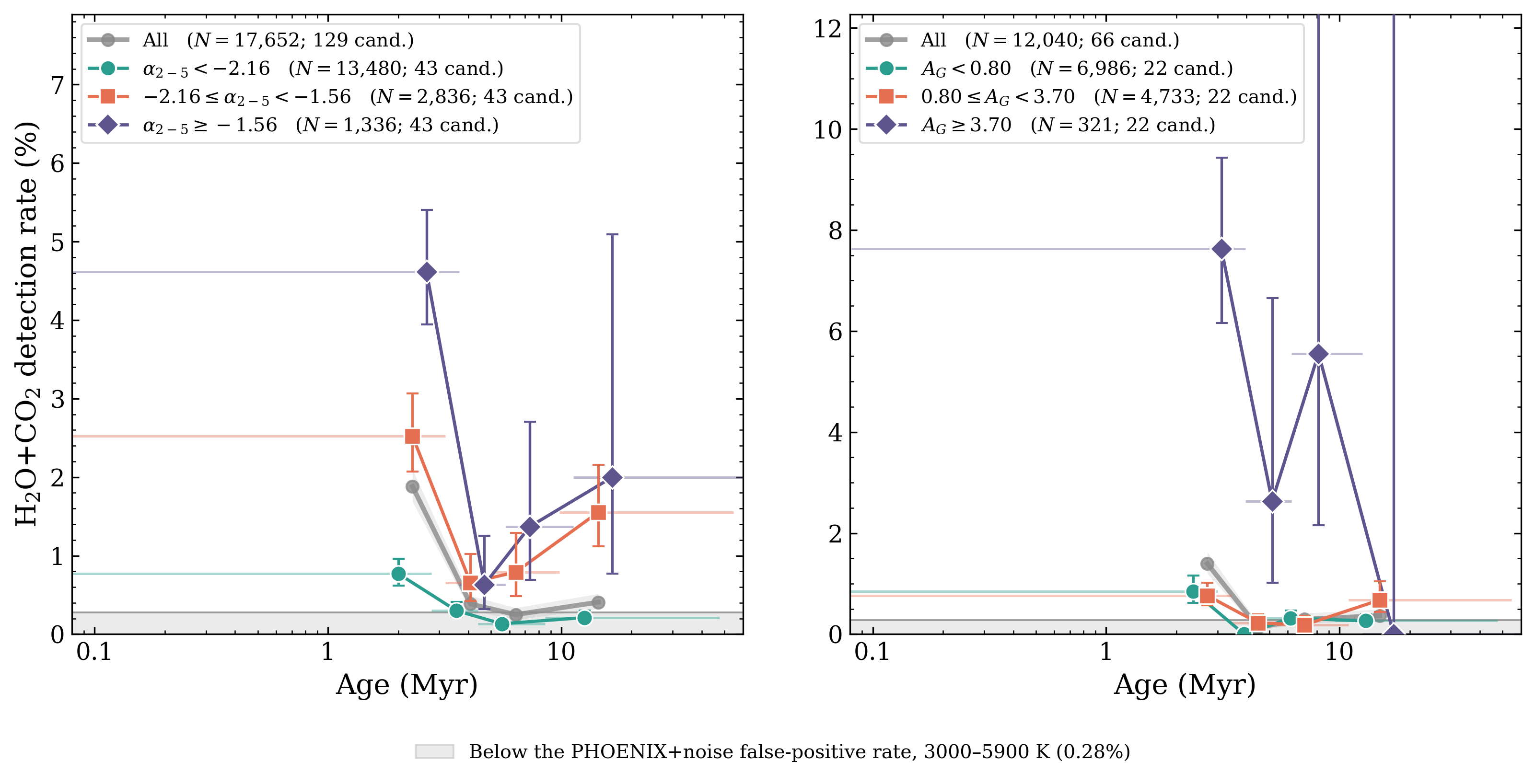}
    \caption{
        H$_2$O+CO$_2$ detection rate versus age within bounded strata of SED slope $\alpha_{2-5\,\mu\rm m}$ (left) and extinction $A_G$ (right). Strata edges are the terciles of the candidates' values, so each stratum holds one third of the candidates; the legend gives each stratum's range, number of evaluable sources, and number of candidates. All strata share the same four equal-population age bins. Error bars are $68\%$ Wilson confidence intervals; horizontal bars show the range of each age bin. The soft gray line and band show the detection rate of all sources in each panel, in the same age bins. The gray band extends up to the contamination level of Figure~\ref{fig:demographics} ($0.28\%$).
    }
    \label{fig:ag_robustness}
\end{figure*}

Younger stars in our sample tend to have steeper near-infrared SED slopes and higher extinction, both of which independently favor detection. To test whether the age trend of Section~\ref{sec:demographics} survives at fixed values of these variables, we split the jointly evaluable sample into three bounded bins of each control variable, with edges at the terciles of the class~A candidates' values, so that each stratum contains the same number of candidates and only the outer strata are open-ended. Within each stratum we compute the detection rate in four age bins of equal population, the same for every stratum, so that the curves are directly comparable (Figure~\ref{fig:ag_robustness}).

\textit{SED slope.} The strata are $\alpha_{2-5\,\mu\rm m}<-2.16$, $-2.16\le\alpha_{2-5\,\mu\rm m}<-1.56$, and $\alpha_{2-5\,\mu\rm m}\ge-1.56$ (43 candidates each). In all three the youngest age bin ($\lesssim3.2$~Myr) exceeds the contamination level by factors of 2.8, 9.0, and 16.5 (Poisson $p\le7\times10^{-5}$). In the flattest stratum the older bins are consistent with contamination; in the two steeper strata they remain $\sim$3.5 times above it, so steep-SED sources retain some detectable ice at older ages. A logistic regression of detection on $\log({\rm age})$ gives $-0.84\pm0.22$ per dex ($p=2\times10^{-4}$) with slope-stratum fixed effects, and $-0.81\pm0.22$ with the slope as a continuous covariate. The age trend is therefore not driven by the age--SED slope correlation.

\textit{Extinction.} $A_G$ is available for 66 of the candidates (22 per stratum: $A_G<0.80$, $0.80\le A_G<3.70$, and $A_G\ge3.70$~mag). In all three strata the youngest age bin ($\lesssim3.5$~Myr) exceeds the contamination level, by factors of 3.0, 2.7, and 27 ($p\approx2\times10^{-3}$ in the two lower strata), and in the two lower strata the older bins are consistent with contamination. The highest stratum contains only 59 sources older than 3.5~Myr, too few to test an age dependence. A model linear in $\log({\rm age})$ with $A_G$-stratum fixed effects does not detect a significant trend ($-0.36\pm0.35$ per dex, $p=0.30$), because the decline is concentrated at young ages rather than being linear in $\log({\rm age})$. A step model separating ages $\lesssim3.5$~Myr, adopted after inspecting Figure~\ref{fig:demographics}c and therefore not an independent test, gives an odds ratio of 3.1 ($p=1\times10^{-4}$) with stratum fixed effects, and of 2.0 ($p=0.03$) when $A_G$ and SED slope enter as continuous covariates. With the present sample, the age dependence at fixed extinction is supported but less firmly established than at fixed SED slope; future SPHEREx cycles and a larger Gaia-extincted subsample can test it directly.

%\newpage

\clearpage

\bibliographystyle{aasjournalv7}
\bibliography{main}

@article{Kounkel_2019b,
doi = {10.3847/1538-3881/ab339a},
url = {https://doi.org/10.3847/1538-3881/ab339a},
year = {2019},
month = {aug},
publisher = {The American Astronomical Society},
volume = {158},
number = {3},
pages = {122},
author = {Kounkel, Marina and Covey, Kevin},
title = {Untangling the Galaxy. I. Local Structure and Star Formation History of the Milky Way},
journal = {The Astronomical Journal}
}

@Article{Hunter2007,
  Author    = {Hunter, J. D.},
  Title     = {Matplotlib: A 2D graphics environment},
  Journal   = {Computing in Science \& Engineering},
  Volume    = {9},
  Number    = {3},
  Pages     = {90--95},
  publisher = {IEEE COMPUTER SOC},
  doi       = {10.1109/MCSE.2007.55},
  year      = 2007
}

@ARTICLE{2013A&A...558A..33A,
       author = {{Astropy Collaboration} and {Robitaille}, Thomas P. and
         {Tollerud}, Erik J. and {Greenfield}, Perry and {Droettboom}, Michael and
         {Bray}, Erik and {Aldcroft}, Tom and {Davis}, Matt and
         {Ginsburg}, Adam and {Price-Whelan}, Adrian M. and
         {Kerzendorf}, Wolfgang E. and {Conley}, Alexander and {Crighton}, Neil and
         {Barbary}, Kyle and {Muna}, Demitri and {Ferguson}, Henry and
         {Grollier}, Fr{\'e}d{\'e}ric and {Parikh}, Madhura M. and
         {Nair}, Prasanth H. and {Unther}, Hans M. and {Deil}, Christoph and
         {Woillez}, Julien and {Conseil}, Simon and {Kramer}, Roban and
         {Turner}, James E.~H. and {Singer}, Leo and {Fox}, Ryan and
         {Weaver}, Benjamin A. and {Zabalza}, Victor and {Edwards}, Zachary I. and
         {Azalee Bostroem}, K. and {Burke}, D.~J. and {Casey}, Andrew R. and
         {Crawford}, Steven M. and {Dencheva}, Nadia and {Ely}, Justin and
         {Jenness}, Tim and {Labrie}, Kathleen and {Lim}, Pey Lian and
         {Pierfederici}, Francesco and {Pontzen}, Andrew and {Ptak}, Andy and
         {Refsdal}, Brian and {Servillat}, Mathieu and {Streicher}, Ole},
        title = "{Astropy: A community Python package for astronomy}",
      journal = {\aap},
         year = "2013",
        month = "Oct",
       volume = {558},
          eid = {A33},
        pages = {A33},
          doi = {10.1051/0004-6361/201322068},
archivePrefix = {arXiv},
       eprint = {1307.6212},
 primaryClass = {astro-ph.IM},
       adsurl = {https://ui.adsabs.harvard.edu/abs/2013A&A...558A..33A}
}

@Article{Harris2020,
 title         = {Array programming with {NumPy}},
 author        = {Charles R. Harris and K. Jarrod Millman and St{'{e}}fan J.
                 van der Walt and Ralf Gommers and Pauli Virtanen and David
                 Cournapeau and Eric Wieser and Julian Taylor and Sebastian
                 Berg and Nathaniel J. Smith and Robert Kern and Matti Picus
                 and Stephan Hoyer and Marten H. van Kerkwijk and Matthew
                 Brett and Allan Haldane and Jaime Fern{'{a}}ndez del
                 R{'{\i}}o and Mark Wiebe and Pearu Peterson and Pierre
                 G{'{e}}rard-Marchant and Kevin Sheppard and Tyler Reddy and
                 Warren Weckesser and Hameer Abbasi and Christoph Gohlke and
                 Travis E. Oliphant},
 year          = {2020},
 month         = sep,
 journal       = {Nature},
 volume        = {585},
 number        = {7825},
 pages         = {357--362},
 doi           = {10.1038/s41586-020-2649-2},
 publisher     = {Springer Science and Business Media {LLC}},
 url           = {https://doi.org/10.1038/s41586-020-2649-2}
}

@ARTICLE{Bock_2025,
       author = {{Bock}, James J. and {Aboobaker}, Asad M. and {Adamo}, Joseph and {Akeson}, Rachel and {Alred}, John M. and {Alibay}, Farah and {Ashby}, Matthew L.~N. and {Bach}, Yoonsoo P. and {Bleem}, Lindsey E. and {Bolton}, Douglas and {Braun}, David F. and {Bruton}, Sean and {Bryan}, Sean A. and {Chang}, Tzu-Ching and {Chen}, Shuang-Shuang and {Cheng}, Yun-Ting and {Cheshire}, IV, James R. and {Chiang}, Yi-Kuan and {Choppin de Janvry}, Jean and {Condon}, Samuel and {Cook}, Walter R. and {Cooray}, Asantha and {Crill}, Brendan P. and {Cukierman}, Ari J. and {Dore}, Olivier and {Dowell}, C. Darren and {Dubois-Felsmann}, Gregory P. and {Eifler}, Tim and {Everett}, Spencer and {Fabinsky}, Beth E. and {Faisst}, Andreas L. and {Fanson}, James L. and {Farrington}, Allen H. and {Fatahi}, Tamim and {Fazar}, Candice M. and {Feder}, Richard M. and {Frater}, Eric H. and {Grasshorn Gebhardt}, Henry S. and {Giri}, Utkarsh and {Goldina}, Tatiana and {Gorjian}, Varoujan and {Habib}, Salman and {Hart}, William G. and {Heinrich}, Chen and {Hora}, Joseph L. and {Huai}, Zhaoyu and {Hui}, Howard and {Jo}, Young-Soo and {Jeong}, Woong-Seob and {Kang}, Jae Hwan and {Kang}, Miju and {Kecman}, Branislav and {Kim}, Chul-Hwan and {Kim}, Jaeyeong and {Kim}, Minjin and {Kim}, Young-Jun and {Kim}, Yongjung and {Kirkpatrick}, J. Davy and {Kobayashi}, Yosuke and {Korngut}, Phil M. and {Krause}, Elisabeth and {Lee}, Bomee and {Lee}, Ho-Gyu and {Lee}, Jae-Joon and {Lee}, Jeong-Eun and {Lisse}, Carey M. and {Mariani}, Giacomo and {Masters}, Daniel C. and {Mauskopf}, Philip D. and {Melnick}, Gary J. and {Minasyan}, Mary H. and {Mirocha}, Jordan and {Miyasaka}, Hiromasa and {Moore}, Anne and {Moore}, Bradley D. and {Murgia}, Giulia and {Naylor}, Bret J. and {Nelson}, Christina and {Nguyen}, Chi H. and {Nguyen}, Hien T. and {Noh}, Jinyoung K. and {Padin}, Stephen and {Paladini}, Roberta and {Park}, Sung-Joon and {Penanen}, Konstantin I. and {Putnam}, Dustin S. and {Pyo}, Jeonghyun and {Ramachandra}, Nesar and {Ramanathan}, Keshav and {Rustamkulov}, Zafar and {Reiley}, Daniel J. and {Rice}, Eric B. and {Rocca}, Jennifer M. and {Seok}, Ji Yeon and {Smith}, Roger and {Stober}, Jeremy and {Susca}, Sara and {Teplitz}, Harry I. and {Thelen}, Michael P. and {Tolls}, Volker and {Torrini}, Gabriela and {Trangsrud}, Amy R. and {Unwin}, Stephen and {Velicheti}, Phani and {Wang}, Pao-Yu and {Wen}, Robin Y. and {-Werner}, Michael-W. and {Williams}, Abby E. and {Williamson}, Ross and {Wincentsen}, James and {Windhorst}, Rogier A. and {Yang}, Soung-Chul and {Yang}, Yujin and {Zemcov}, Michael},
        title = "{The SPHEREx Satellite Mission}",
      journal = {arXiv e-prints},
         year = 2025,
        month = nov,
          eid = {arXiv:2511.02985},
        pages = {arXiv:2511.02985},
          doi = {10.48550/arXiv.2511.02985},
archivePrefix = {arXiv},
       eprint = {2511.02985},
 primaryClass = {astro-ph.IM},
       adsurl = {https://ui.adsabs.harvard.edu/abs/2025arXiv251102985B}
}

@ARTICLE{Akeson_2025,
       author = {{Akeson}, Rachel and {Dubois-Felsmann}, Gregory P. and {Crill}, Brendan P. and {Faisst}, Andreas L. and {Fatahi}, Tamim and {Fazar}, Candice M. and {Goldina}, Tatiana and {Masters}, Daniel C. and {Nelson}, Christina and {Paladini}, Roberta and {Teplitz}, Harry I. and {Torrini}, Gabriela and {Velicheti}, Phani and {Ashby}, Matthew L.~N. and {Avner}, Dan and {Bach}, Yoonsoo P. and {Bock}, James J. and {Bruton}, Sean and {Bryan}, Sean A. and {Chang}, Tzu-Ching and {Chen}, Shuang-Shuang and {Cukierman}, Ari J. and {Dore}, O. and {Dowell}, C. Darren and {Everett}, Spencer and {Feder}, Richard M. and {Huai}, Zhaoyu and {Hui}, Howard and {Jeong}, Woong-Seob and {Jo}, Young-Soo and {Korngut}, Phil M. and {Kwon}, Yuna G. and {Lee}, Bomee and {Melnick}, Gary J. and {Murgia}, Giulia and {Nguyen}, Chi H. and {Pourrahmani}, Milad and {Rustamkulov}, Zafar and {Tolls}, Volker and {Wang}, Pao-Yu and {Yang}, Yujin and {Zemcov}, Michael},
        title = "{The SPHEREx Image and Spectrophotometry Processing Pipeline}",
      journal = {arXiv e-prints},
         year = 2025,
        month = nov,
          eid = {arXiv:2511.15823},
        pages = {arXiv:2511.15823},
          doi = {10.48550/arXiv.2511.15823},
archivePrefix = {arXiv},
       eprint = {2511.15823},
 primaryClass = {astro-ph.IM},
       adsurl = {https://ui.adsabs.harvard.edu/abs/2025arXiv251115823A}
}

@article{Gibb_2004,
doi = {10.1086/381182},
url = {https://doi.org/10.1086/381182},
year = {2004},
month = {mar},
publisher = {},
volume = {151},
number = {1},
pages = {35},
author = {Gibb, E. L. and Whittet, D. C. B. and Boogert, A. C. A. and Tielens, A. G. G. M.},
title = {Interstellar Ice: The Infrared Space Observatory Legacy},
journal = {The Astrophysical Journal Supplement Series}
}

@article{Pontoppidan_2008,
doi = {10.1086/533431},
url = {https://doi.org/10.1086/533431},
year = {2008},
month = {may},
publisher = {},
volume = {678},
number = {2},
pages = {1005},
author = {Pontoppidan, Klaus M. and Boogert, A. C. A. and Fraser, Helen J. and van Dishoeck, Ewine F. and Blake, Geoffrey A. and Lahuis, Fred and Öberg, Karin I. and Evans II, Neal J. and Salyk, Colette},
title = {The c2d Spitzer Spectroscopic Survey of Ices around Low-Mass Young Stellar Objects. II. CO2},
journal = {The Astrophysical Journal}
}

@article{Boogert_2008,
doi = {10.1086/533425},
url = {https://doi.org/10.1086/533425},
year = {2008},
month = {may},
publisher = {},
volume = {678},
number = {2},
pages = {985},
author = {Boogert, A. C. A. and Pontoppidan, K. M. and Knez, C. and Lahuis, F. and Kessler-Silacci, J. and van Dishoeck, E. F. and Blake, G. A. and Augereau, J.-C. and Bisschop, S. E. and Bottinelli, S. and Brooke, T. Y. and Brown, J. and Crapsi, A. and Evans II, N. J. and Fraser, H. J. and Geers, V. and Huard, T. L. and Jørgensen, J. K. and Öberg, K. I. and Allen, L. E. and Harvey, P. M. and Koerner, D. W. and Mundy, L. G. and Padgett, D. L. and Sargent, A. I. and Stapelfeldt, K. R.},
title = {The c2d Spitzer Spectroscopic Survey of Ices around Low-Mass Young Stellar Objects. I. H2O and the 5-8 μm Bands* **},
journal = {The Astrophysical Journal}
}

@article{Aikawa_2012,
	author = {{Aikawa, Y.} and {Kamuro, D.} and {Sakon, I.} and {Itoh, Y.} and {Terada, H.} and {Noble, J. A.} and {Pontoppidan, K. M.} and {Fraser, H. J.} and {Tamura, M.} and {Kandori, R.} and {Kawamura, A.} and {Ueno, M.}},
	title = {AKARI observations of ice absorption bands towards edge-on young stellar objects},
	DOI= "10.1051/0004-6361/201015999",
	url= "https://doi.org/10.1051/0004-6361/201015999",
	journal = {A\&A},
	year = 2012,
	volume = 538,
	pages = "A57",
	month = "",
}

@ARTICLE{McClure_2025,
       author = {{McClure}, M.~K. and {van't Hoff}, Merel and {Francis}, Logan and {Bergin}, Edwin and {Rocha}, Will R.~M. and {Sturm}, J.~A. and {Harsono}, Daniel and {van Dishoeck}, Ewine F. and {Black}, John H. and {Noble}, J.~A. and {Qasim}, D. and {Dartois}, E.},
        title = "{Refractory solid condensation detected in an embedded protoplanetary disk}",
      journal = {\nat},
         year = 2025,
        month = jul,
       volume = {643},
       number = {8072},
        pages = {649-653},
          doi = {10.1038/s41586-025-09163-z},
       adsurl = {https://ui.adsabs.harvard.edu/abs/2025Natur.643..649M}
}

@ARTICLE{Kim_2025,
       author = {{Kim}, Jaeyeong and {Lee}, Jeong-Eun and {Kim}, Chul-Hwan and {Jeong}, Woong-Seob and {Yang}, Yao-Lun},
        title = "{Near- to Mid-Infrared Spectroscopic Study of Ice Analysis Using the AKARI/IRC and Spitzer/IRS Spectra}",
      journal = {Journal of Korean Astronomical Society},
         year = 2025,
        month = may,
       volume = {58},
        pages = {111-129},
          doi = {10.5303/JKAS.2025.58.1.111},
archivePrefix = {arXiv},
       eprint = {2505.05390},
 primaryClass = {astro-ph.SR},
       adsurl = {https://ui.adsabs.harvard.edu/abs/2025JKAS...58..111K}
}

@article{Dartois_2025,
	author = {{Dartois, E.} and {Noble, J. A.} and {McClure, M. K.} and {Sturm, J. A.} and {Beck, T. L.} and {Arulanantham, N.} and {Drozdovskaya, M. N.} and {Espaillat, C. C.} and {Harsono, D.} and {Palumbo, M.-E.} and {Pendleton, Y. J.} and {Pontoppidan, K. M.}},
	title = {The edge-on disc Tau 042021: Icy grains at high altitudes and a wind containing astronomical polycyclic aromatic hydrocarbons},
	DOI= "10.1051/0004-6361/202452966",
	url= "https://doi.org/10.1051/0004-6361/202452966",
	journal = {A\&A},
	year = 2025,
	volume = 698,
	pages = "A8",
}

@ARTICLE{Smith_2025,
       author = {{Smith}, Z.~L. and {Dickinson}, H.~J. and {Fraser}, H.~J. and {McClure}, M.~K. and {Noble}, J.~A. and {Boogert}, A.~C.~A. and {Sun}, F. and {Egami}, E. and {Dartois}, E. and {Erkal}, J. and {Shimonishi}, T. and {Beck}, T.~L. and {Bergner}, J.~B. and {Caselli}, P. and {Charnley}, S.~B. and {Chu}, L. and {Drozdovskaya}, M.~N. and {Garrod}, R. and {Harsono}, D. and {Ioppolo}, S. and {Jimenez-Serra}, I. and {J{\o}rgensen}, J.~K. and {Melnick}, G.~J. and {{\~A}-berg}, K.~I. and {Palumbo}, M.~E. and {Pendleton}, Y.~J. and {Perotti}, G. and {Pontoppidan}, K.~M. and {Qasim}, D. and {Rocha}, W.~R.~M. and {Sturm}, J.~A. and {Taillard}, A. and {Urso}, R.~G. and {van Dishoeck}, E.~F.},
        title = "{Cospatial ice mapping of H$_{2}$O with CO$_{2}$ and CO across a molecular cloud with JWST/NIRCam}",
      journal = {Nature Astronomy},
         year = 2025,
        month = jun,
       volume = {9},
        pages = {883-894},
          doi = {10.1038/s41550-025-02511-z},
       adsurl = {https://ui.adsabs.harvard.edu/abs/2025NatAs...9..883S}
}

@article{Dartois_2026,
	author = {{Dartois, E.} and {Noble, Jennifer A.} and {Bergner, Jennifer B.} and {Pontoppidan, Klaus M.} and {Assani, Korash} and {Harsono, Daniel} and {McClure, Melissa K.} and {Santos, Julia C.} and {Thompson, Will E.} and {Welzel, Lukas} and {Arulanantham, Nicole} and {Booth, Alice S.} and {Drozdovskaya, Maria N.} and {Li, Zhi-Yun} and {Ma, Jie} and {Martinien, Laurine} and {Ménard, François} and {Oberg, Karin} and {Stapelfeldt, Karl} and {Yang, Yao-Lun}},
	title = {The near-infrared silhouette of the Flying Saucer edge-on disc revealed by the JWST JEDIce program},
	DOI= "10.1051/0004-6361/202557177",
	url= "https://doi.org/10.1051/0004-6361/202557177",
	journal = {A\&A},
	year = 2026,
	volume = 705,
	pages = "L14",
}

@article{Delabrosse_2024,
	author = {{Delabrosse, V.} and {Dougados, C.} and {Cabrit, S.} and {Tabone, B.} and {Tychoniec, L.} and {Ray, T.} and {Podio, L.} and {McClure, M.}},
	title = {JWST study of the DG Tau B disk-wind candidate - I. Overview and nested H2-CO outflows},
	DOI= "10.1051/0004-6361/202449176",
	url= "https://doi.org/10.1051/0004-6361/202449176",
	journal = {A\&A},
	year = 2024,
	volume = 688,
	pages = "A173",
}

@article{Slavicinska_2024,
	author = {{Slavicinska, Katerina} and {van Dishoeck, Ewine F.} and {Tychoniec, Łukasz} and {Nazari, Pooneh} and {Rubinstein, Adam E.} and {Gutermuth, Robert} and {Tyagi, Himanshu} and {Chen, Yuan} and {Brunken, Nashanty G. C.} and {Rocha, Will R. M.} and {Manoj, P.} and {Narang, Mayank} and {Thomas Megeath, S.} and {Yang, Yao-Lun} and {Looney, Leslie W.} and {Tobin, John J.} and {Beuther, Henrik} and {Bourke, Tyler L.} and {Linnartz, Harold} and {Federman, Samuel} and {Watson, Dan M.} and {Linz, Hendrik}},
	title = {JWST detections of amorphous and crystalline HDO ice toward massive protostars},
	DOI= "10.1051/0004-6361/202449785",
	url= "https://doi.org/10.1051/0004-6361/202449785",
	journal = {A\&A},
	year = 2024,
	volume = 688,
	pages = "A29",
}

@article{Arulanantham_2024,
doi = {10.3847/2041-8213/ad35c9},
url = {https://doi.org/10.3847/2041-8213/ad35c9},
year = {2024},
month = {apr},
publisher = {The American Astronomical Society},
volume = {965},
number = {1},
pages = {L13},
author = {Arulanantham, Nicole and McClure, M. K. and Pontoppidan, Klaus and Beck, Tracy L. and Sturm, J. A. and Harsono, D. and Boogert, A. C. A. and Cordiner, M. and Dartois, E. and Drozdovskaya, M. N. and Espaillat, C. and Melnick, G. J. and Noble, J. A. and Palumbo, M. E. and Pendleton, Y. J. and Terada, H. and van Dishoeck, E. F.},
title = {JWST MIRI MRS Images of Disk Winds, Water, and CO in an Edge-on Protoplanetary Disk},
journal = {The Astrophysical Journal Letters}
}

@article{Boogert_2015,
   author = "Boogert, A.C. Adwin and Gerakines, Perry A. and Whittet, Douglas C.B.",
   title = "Observations of the Icy Universe", 
   journal= "Annual Review of Astronomy and Astrophysics",
   year = "2015",
   volume = "53",
   number = "Volume 53, 2015",
   pages = "541-581",
   doi = "https://doi.org/10.1146/annurev-astro-082214-122348",
   url = "https://www.annualreviews.org/content/journals/10.1146/annurev-astro-082214-122348",
   publisher = "Annual Reviews",
   issn = "1545-4282",
   type = "Journal Article",
  }

@article{Oberg_2007,
	author = {{Öberg, K. I.} and {Fraser, H. J.} and {Boogert, A. C. A.} and {Bisschop, S. E.} and {Fuchs, G. W.} and {van Dishoeck, E. F.} and {Linnartz, H.}},
	title = {Effects of CO2 on H2O band profiles and band  strengths in mixed H2O:CO2 ices},
	DOI= "10.1051/0004-6361:20065881",
	url= "https://doi.org/10.1051/0004-6361:20065881",
	journal = {A\&A},
	year = 2007,
	volume = 462,
	number = 3,
	pages = "1187-1198",
}

@article{VanBroekhuizen_2006,
	author = {{van Broekhuizen, F. A.} and {Groot, I. M. N.} and {Fraser, H. J.} and {van Dishoeck, E. F.} and {Schlemmer, S.}},
	title = {Infrared spectroscopy of solid CO–CO2 mixtures and layers},
	DOI= "10.1051/0004-6361:20052942",
	url= "https://doi.org/10.1051/0004-6361:20052942",
	journal = {A\&A},
	year = 2006,
	volume = 451,
	number = 2,
	pages = "723-731",
}

@article{Rocha_2022,
	author = {{Rocha, W. R. M.} and {Rachid, M. G.} and {Olsthoorn, B.} and {van Dishoeck, E. F.} and {McClure, M. K.} and {Linnartz, H.}},
	title = {LIDA: The Leiden Ice Database for Astrochemistry},
	DOI= "10.1051/0004-6361/202244032",
	url= "https://doi.org/10.1051/0004-6361/202244032",
	journal = {A\&A},
	year = 2022,
	volume = 668,
	pages = "A63",
}

@article{Briceno_2019,
	doi = {10.3847/1538-3881/aaf79b},
	url = {https://doi.org/10.3847%2F1538-3881%2Faaf79b},
	year = 2019,
	month = {jan},
	publisher = {American Astronomical Society},
	volume = {157},
	number = {2},
	pages = {85},
	author = {C{\'{e}}sar Brice{\~{n}}o and Nuria Calvet and Jes{\'{u}}s Hern{\'{a}}ndez and A. Katherina Vivas and Cecilia Mateu and Juan Jos{\'{e}} Downes and Jaqueline Loerincs and Alice P{\'{e}}rez-Blanco and Perry Berlind and Catherine Espaillat and Lori Allen and Lee Hartmann and Mario Mateo and John I. Bailey III},
	title = {The {CIDA} Variability Survey of Orion {OB}1. {II}. Demographics of the Young, Low-mass Stellar Populations},
	journal = {The Astronomical Journal}
}

@article{Hernandez_2014,
	doi = {10.1088/0004-637x/794/1/36},
	url = {https://doi.org/10.1088%2F0004-637x%2F794%2F1%2F36},
	year = 2014,
	month = {sep},
	publisher = {{IOP} Publishing},
	volume = {794},
	number = {1},
	pages = {36},
	author = {Jes{\'{u}}s Hern{\'{a}}ndez and Nuria Calvet and Alice Perez and Cesar Brice{\~{n}}o and Lorenzo Olguin and Maria E. Contreras and Lee Hartmann and Lori Allen and Catherine Espaillat and Ram{\'{\i}}rez Hernan},
	title = {A {SPECTROSCOPIC} {CENSUS} {IN} {YOUNG} {STELLAR} {REGIONS}: {THE} $\upsigma$ {ORIONIS} {CLUSTER}},
	journal = {The Astrophysical Journal}
}

@article{Dishoeck_2025,
	author = {{van Dishoeck, E. F.} and {Tychoniec, Ł.} and {Rocha, W. R. M.} and {Slavicinska, K.} and {Francis, L.} and {van Gelder, M. L.} and {Ray, T. P.} and {Beuther, H.} and {Caratti o Garatti, A.} and {Brunken, N. G. C.} and {Chen, Y.} and {Devaraj, R.} and {Geers, V. C.} and {Gieser, C.} and {Greene, T. P.} and {Justtanont, K.} and {Le Gouellec, V. J. M.} and {Kavanagh, P. J.} and {Klaassen, P. D.} and {Janssen, A. G. M.} and {Navarro, M. G.} and {Nazari, P.} and {Notsu, S.} and {Perotti, G.} and {Ressler, M. E.} and {Reyes, S. D.} and {Sellek, A. D.} and {Tabone, B.} and {Tap, C.} and {Theijssen, N. C. M. A.} and {Colina, L.} and {Güdel, M.} and {Henning, Th.} and {Lagage, P.-O.} and {Östlin, G.} and {Vandenbussche, B.} and {Wright, G. S.}},
	title = {JWST Observations of Young protoStars (JOYS) - Overview of program and early results},
	DOI= "10.1051/0004-6361/202554444",
	url= "https://doi.org/10.1051/0004-6361/202554444",
	journal = {A\&A},
	year = 2025,
	volume = 699,
	pages = "A361",
}

@ARTICLE{Dartois_2024,
       author = {{Dartois}, E. and {Noble}, J.~A. and {Caselli}, P. and {Fraser}, H.~J. and {Jim{\'e}nez-Serra}, I. and {Mat{\'e}}, B. and {McClure}, M.~K. and {Melnick}, G.~J. and {Pendleton}, Y.~J. and {Shimonishi}, T. and {Smith}, Z.~L. and {Sturm}, J.~A. and {Taillard}, A. and {Wakelam}, V. and {Boogert}, A.~C.~A. and {Drozdovskaya}, M.~N. and {Erkal}, J. and {Harsono}, D. and {Herrero}, V.~J. and {Ioppolo}, S. and {Linnartz}, H. and {McGuire}, B.~A. and {Perotti}, G. and {Qasim}, D. and {Rocha}, W.~R.~M.},
        title = "{Spectroscopic sizing of interstellar icy grains with JWST}",
      journal = {Nature Astronomy},
         year = 2024,
        month = mar,
       volume = {8},
        pages = {359-367},
          doi = {10.1038/s41550-023-02155-x},
       adsurl = {https://ui.adsabs.harvard.edu/abs/2024NatAs...8..359D}
}

@ARTICLE{McClure_2023,
       author = {{McClure}, M.~K. and {Rocha}, W.~R.~M. and {Pontoppidan}, K.~M. and {Crouzet}, N. and {Chu}, L.~E.~U. and {Dartois}, E. and {Lamberts}, T. and {Noble}, J.~A. and {Pendleton}, Y.~J. and {Perotti}, G. and {Qasim}, D. and {Rachid}, M.~G. and {Smith}, Z.~L. and {Sun}, Fengwu and {Beck}, Tracy L. and {Boogert}, A.~C.~A. and {Brown}, W.~A. and {Caselli}, P. and {Charnley}, S.~B. and {Cuppen}, Herma M. and {Dickinson}, H. and {Drozdovskaya}, M.~N. and {Egami}, E. and {Erkal}, J. and {Fraser}, H. and {Garrod}, R.~T. and {Harsono}, D. and {Ioppolo}, S. and {Jim{\'e}nez-Serra}, I. and {Jin}, M. and {J{\o}rgensen}, J.~K. and {Kristensen}, L.~E. and {Lis}, D.~C. and {McCoustra}, M.~R.~S. and {McGuire}, Brett A. and {Melnick}, G.~J. and {{\"O}berg}, Karin I. and {Palumbo}, M.~E. and {Shimonishi}, T. and {Sturm}, J.~A. and {van Dishoeck}, E.~F. and {Linnartz}, H.},
        title = "{An Ice Age JWST inventory of dense molecular cloud ices}",
      journal = {Nature Astronomy},
         year = 2023,
        month = apr,
       volume = {7},
        pages = {431-443},
          doi = {10.1038/s41550-022-01875-w},
archivePrefix = {arXiv},
       eprint = {2301.09140},
 primaryClass = {astro-ph.GA},
       adsurl = {https://ui.adsabs.harvard.edu/abs/2023NatAs...7..431M}
}

@article{Sturm_2023,
	author = {{Sturm, J. A.} and {McClure, M. K.} and {Beck, T. L.} and {Harsono, D.} and {Bergner, J. B.} and {Dartois, E.} and {Boogert, A. C. A.} and {Chiar, J. E.} and {Cordiner, M. A.} and {Drozdovskaya, M. N.} and {Ioppolo, S.} and {Law, C. J.} and {Linnartz, H.} and {Lis, D. C.} and {Melnick, G. J.} and {McGuire, B. A.} and {Noble, J. A.} and {Öberg, K. I.} and {Palumbo, M. E.} and {Pendleton, Y. J.} and {Perotti, G.} and {Pontoppidan, K. M.} and {Qasim, D.} and {Rocha, W. R. M.} and {Terada, H.} and {Urso, R. G.} and {van Dishoeck, E. F.}},
	title = {A JWST inventory of protoplanetary disk ices - The edge-on protoplanetary disk HH 48 NE, seen with the Ice Age ERS program★},
	DOI= "10.1051/0004-6361/202347512",
	url= "https://doi.org/10.1051/0004-6361/202347512",
	journal = {A\&A},
	year = 2023,
	volume = 679,
	pages = "A138",
}

@article{Ashby_2023,
doi = {10.3847/1538-4357/acc86b},
url = {https://doi.org/10.3847/1538-4357/acc86b},
year = {2023},
month = {jun},
publisher = {The American Astronomical Society},
volume = {949},
number = {2},
pages = {105},
author = {Ashby, Matthew L. N. and Hora, Joseph L. and Lakshmipathaiah, Kiran and Vig, Sarita and Sai Subrahmanyam Gorthi, Rama Krishna and Kang, Miju and Tolls, Volker and Melnick, Gary J. and Werner, Michael W. and Crill, Brendan P. and Masters, Daniel C. and Peña, Carlos Contreras and Lee, Jeong-Eun and Kim, Jaeyeong and Lee, Ho-Gyu and Yoon, Sung-Yong and Yang, Soung-Chul and Flagey, Nicholas and Mennesson, Bertrand},
title = {The SPHEREx Target List of Ice Sources (SPLICES)},
journal = {The Astrophysical Journal}
}

@article{Allard_2012,
  author  = {Allard, F. and Homeier, D. and Freytag, B.},
  title   = {Models of very-low-mass stars, brown dwarfs and exoplanets},
  journal = {Philosophical Transactions of the Royal Society A},
  volume  = {370},
  number  = {1968},
  pages   = {2765--2777},
  year    = {2012},
  doi     = {10.1098/rsta.2011.0269}
}

@article{Asplund_2009,
  author  = {Asplund, M. and Grevesse, N. and Sauval, A. J. and Scott, P.},
  title   = {The Chemical Composition of the Sun},
  journal = {Annual Review of Astronomy and Astrophysics},
  volume  = {47},
  pages   = {481--522},
  year    = {2009},
  doi     = {10.1146/annurev.astro.46.060407.145222}
}

@article{Barber_2006,
  author  = {Barber, R. J. and Tennyson, J. and Harris, G. J. and Tolchenov, R. N.},
  title   = {A high-accuracy computed water line list},
  journal = {Monthly Notices of the Royal Astronomical Society},
  volume  = {368},
  pages   = {1087--1094},
  year    = {2006},
  doi     = {10.1111/j.1365-2966.2006.10184.x}
}

@article{Youden_1950,
author = {Youden, W. J.},
title = {Index for rating diagnostic tests},
journal = {Cancer},
volume = {3},
number = {1},
pages = {32-35},
doi = {https://doi.org/10.1002/1097-0142(1950)3:1<32::AID-CNCR2820030106>3.0.CO;2-3},
url = {https://acsjournals.onlinelibrary.wiley.com/doi/abs/10.1002/1097-0142%281950%293%3A1%3C32%3A%3AAID-CNCR2820030106%3E3.0.CO%3B2-3},
eprint = {https://acsjournals.onlinelibrary.wiley.com/doi/pdf/10.1002/1097-0142%281950%293%3A1%3C32%3A%3AAID-CNCR2820030106%3E3.0.CO%3B2-3},
year = {1950}
}

@article{Melnick_2026,
doi = {10.3847/1538-4357/ae53e0},
url = {https://doi.org/10.3847/1538-4357/ae53e0},
year = {2026},
month = {apr},
publisher = {The American Astronomical Society},
volume = {1002},
number = {1},
pages = {33},
author = {Melnick, Gary J. and Hora, Joseph L. and Ashby, Matthew L. N. and Tolls, Volker and Kim, Jaeyeong and Lisse, Carey M. and Paladini, Roberta and Werner, Michael W. and Lee, Jeong-Eun and Kim, Young-Jun and Kang, Miju and Cheng, Yun-Ting and Bock, James J. and Crill, Brendan P. and Cukierman, Ari and Doré, Olivier and Faisst, Andreas and Hui, Howard and Jeong, Woong-Seob and Kim, Chul-Hwan and Lee, Ho-Gyu and Lee, Jae-Joon and Masters, Daniel and Nguyen, Chi H. and Noh, Jinyoung and Seok, Ji Yeon and Yang, Soung-Chul and Yang, Yujin and Zemcov, Michael},
title = {The SPHEREx Ices Investigation: An Overview},
journal = {The Astrophysical Journal}
}

@article{Hora_2026,
doi = {10.3847/1538-4357/ae5180},
url = {https://doi.org/10.3847/1538-4357/ae5180},
year = {2026},
month = {apr},
publisher = {The American Astronomical Society},
volume = {1001},
number = {2},
pages = {165},
author = {Hora, Joseph L. and Noh, Jinyoung K. and Melnick, Gary J. and Hensley, Brandon S. and Paladini, Roberta and Lee, Jeong-Eun and Ashby, Matthew L. N. and Tolls, Volker and Kim, Jaeyeong and Werner, Michael W. and Bock, James J. and Bruton, Sean and Chen, Shuang-Shuang and Chang, Tzu-Ching and Chiang, Yi-Kuan and Cooray, Asantha and Crill, Brendan P. and Cukierman, Ari J. and Doré, Olivier and Faisst, Andreas L. and Huai, Zhaoyu and Hui, Howard and Jeong, Woong-Seob and Kang, Miju and Korngut, Phil M. and Lee, Ho-Gyu and Lisse, Carey M. and Masters, Daniel C. and Murgia, Giulia and Nguyen, Chi H. and Rustamkulov, Zafar and Seok, Ji Yeon and Wen, Robin Y. and Yang, Yujin and Zemcov, Michael},
title = {SPHEREx Widefield Infrared Spectral Mapping of Interstellar Ices and Polycyclic Aromatic Hydrocarbons},
journal = {The Astrophysical Journal}
}

@article{Murgia_2026,
  title={SPHEREx mapping of diffuse PAH and H II emission in the Galactic plane},
  author={Murgia, Giulia and Cukierman, Ari J and Hensley, Brandon S and Ashby, Matthew LN and Bock, James J and Chang, Tzu-Ching and Chen, Shuang-Shuang and Cheng, Yun-Ting and Chiang, Yi-Kuan and Cooray, Asantha and others},
  journal={arXiv preprint arXiv:2603.23620},
  year={2026}
}

@article{McBride_2021,
doi = {10.3847/1538-3881/ac2432},
url = {https://doi.org/10.3847/1538-3881/ac2432},
year = {2021},
month = {dec},
publisher = {The American Astronomical Society},
volume = {162},
number = {6},
pages = {282},
author = {McBride, Aidan and Lingg, Ryan and Kounkel, Marina and Covey, Kevin and Hutchinson, Brian},
title = {Untangling the Galaxy. III. Photometric Search for Pre-main-sequence Stars with Deep Learning},
journal = {The Astronomical Journal}
}

@article{Visser_2009,
	author = {{Visser, R.} and {van Dishoeck, E. F.} and {Doty, S. D.} and {Dullemond, C. P.}},
	title = {The chemical history of molecules in circumstellar disks  - I. Ices },
	DOI= "10.1051/0004-6361/200810846",
	url= "https://doi.org/10.1051/0004-6361/200810846",
	journal = {A\&A},
	year = 2009,
	volume = 495,
	number = 3,
	pages = "881-897",
}

@ARTICLE{Oberg_2021,
       author = {{{\"O}berg}, Karin I. and {Bergin}, Edwin A.},
        title = "{Astrochemistry and compositions of planetary systems}",
      journal = {\physrep},
         year = 2021,
        month = jan,
       volume = {893},
        pages = {1-48},
          doi = {10.1016/j.physrep.2020.09.004},
archivePrefix = {arXiv},
       eprint = {2010.03529},
 primaryClass = {astro-ph.EP},
       adsurl = {https://ui.adsabs.harvard.edu/abs/2021PhR...893....1O}
}

@article{Whitney_2003,
doi = {10.1086/379068},
url = {https://doi.org/10.1086/379068},
year = {2003},
month = {dec},
publisher = {},
volume = {598},
number = {2},
pages = {1079},
author = {Whitney, Barbara A. and Wood, Kenneth and Bjorkman, J. E. and Cohen, Martin},
title = {Two-dimensional Radiative Transfer in Protostellar Envelopes. II. An Evolutionary Sequence},
journal = {The Astrophysical Journal}
}

@article{Eistrup_2016,
	author = {{Eistrup, Christian} and {Walsh, Catherine} and {van Dishoeck, Ewine F.}},
	title = {Setting the volatile composition of (exo)planet-building
          material - Does chemical evolution in disk midplanes matter?},
	DOI= "10.1051/0004-6361/201628509",
	url= "https://doi.org/10.1051/0004-6361/201628509",
	journal = {A\&A},
	year = 2016,
	volume = 595,
	pages = "A83",
}

@article{Williams_2011,
issn = {0066-4146},
journal = {Annual review of astronomy and astrophysics},
pages = {67--117},
volume = {49},
publisher = {Annual Reviews},
number = {1},
year = {2011},
title = {Protoplanetary Disks and Their Evolution},
copyright = {Copyright © 2011 by Annual Reviews. All rights reserved 2011},
language = {eng},
address = {Palo Alto},
author = {Williams, Jonathan P. and Cieza, Lucas A.},
}

@article{Bosman_2018,
	author = {{Bosman, Arthur D.} and {Walsh, Catherine} and {van Dishoeck, Ewine F.}},
	title = {CO destruction in protoplanetary disk midplanes: Inside versus outside the CO snow surface},
	DOI= "10.1051/0004-6361/201833497",
	url= "https://doi.org/10.1051/0004-6361/201833497",
	journal = {A\&A},
	year = 2018,
	volume = 618,
	pages = "A182",
}

@article{Van_Clepper_2022,
doi = {10.3847/1538-4357/ac511b},
url = {https://doi.org/10.3847/1538-4357/ac511b},
year = {2022},
month = {mar},
publisher = {The American Astronomical Society},
volume = {927},
number = {2},
pages = {206},
author = {Van Clepper, Eric and Bergner, Jennifer B. and Bosman, Arthur D. and Bergin, Edwin and Ciesla, Fred J.},
title = {Chemical Feedback of Pebble Growth: Impacts on CO depletion and C/O ratios},
journal = {The Astrophysical Journal}
}

@article{Manzo-Martinez_2020,
doi = {10.3847/1538-4357/ab7ead},
url = {https://doi.org/10.3847/1538-4357/ab7ead},
year = {2020},
month = {apr},
publisher = {The American Astronomical Society},
volume = {893},
number = {1},
pages = {56},
author = {Manzo-Martínez, Ezequiel and Calvet, Nuria and Hernández, Jesús and Lizano, Susana and Hernández, Ramiro Franco and Miller, Christopher J. and Maucó, Karina and Briceño, César and D’Alessio, Paola},
title = {The Evolution of the Inner Regions of Protoplanetary Disks},
journal = {The Astrophysical Journal}
}

@article{Davidsson_2021,
    author = {Davidsson, Björn J R},
    title = {Thermophysical evolution of planetesimals in the primordial disc},
    journal = {Monthly Notices of the Royal Astronomical Society},
    volume = {505},
    number = {4},
    pages = {5654-5685},
    year = {2021},
    month = {08},
    issn = {0035-8711},
    doi = {10.1093/mnras/stab1593},
    url = {https://doi.org/10.1093/mnras/stab1593},
    eprint = {https://academic.oup.com/mnras/article-pdf/505/4/5654/38869740/stab1593.pdf},
}

@article{Sarver_2026,
author = {Sarver, Collette C. and Walker, Catherine E. and Widicus Weaver, Susanna L.},
title = {Laboratory Investigation of Simultaneous Ultraviolet Photoprocessing and Temperature-Programmed Desorption of Interstellar Ice Analogs},
journal = {ACS Earth and Space Chemistry},
volume = {10},
number = {2},
pages = {536-554},
year = {2026},
doi = {10.1021/acsearthspacechem.5c00338},
URL = {https://doi.org/10.1021/acsearthspacechem.5c00338},
eprint = {https://doi.org/10.1021/acsearthspacechem.5c00338}
}

@article{Munoz-Caro_2025,
  author    = {Mu{\~n}oz Caro, Guillermo M. and Carrascosa de Lucas, H{\'e}ctor and Mart{\'i}n-Dom{\'e}nech, Rafael},
  title     = {Photochemistry of interstellar ice forming complex organic molecules},
  journal   = {Nature Reviews Chemistry},
  volume    = {9},
  number    = {8},
  pages     = {537--552},
  year      = {2025},
  month     = aug,
  doi       = {10.1038/s41570-025-00729-z},
  url       = {https://www.nature.com/articles/s41570-025-00729-z},
  publisher = {Nature Publishing Group},
  issn      = {2397-3358}
}

@article{Yu_2017,
doi = {10.3847/1538-4357/aa6e4c},
url = {https://doi.org/10.3847/1538-4357/aa6e4c},
year = {2017},
month = {may},
publisher = {The American Astronomical Society},
volume = {841},
number = {1},
pages = {39},
author = {Yu, Mo and Evans II, Neal J. and Dodson-Robinson, Sarah E. and Willacy, Karen and Turner, Neal J.},
title = {Disk Masses around Solar-mass Stars are Underestimated by CO Observations},
journal = {The Astrophysical Journal}
}

@ARTICLE{Ehrenfreund_1999,
       author = {{Ehrenfreund}, P. and {Kerkhof}, O. and {Schutte}, W.~A. and {Boogert}, A.~C.~A. and {Gerakines}, P.~A. and {Dartois}, E. and {D'Hendecourt}, L. and {Tielens}, A.~G.~G.~M. and {van Dishoeck}, E.~F. and {Whittet}, D.~C.~B.},
        title = "{Laboratory studies of thermally processed H\_2O-CH\_3OH-CO\_2 ice mixtures and their astrophysical implications}",
      journal = {\aap},
         year = 1999,
        month = oct,
       volume = {350},
        pages = {240-253},
       adsurl = {https://ui.adsabs.harvard.edu/abs/1999A&A...350..240E}
}

@article{Goto_2021,
	author = {{Goto, M.} and {Vasyunin, A. I.} and {Giuliano, B. M.} and {Jiménez-Serra, I.} and {Caselli, P.} and {Román-Zúñiga, C. G.} and {Alves, J.}},
	title = {Water and methanol ice in L 1544★},
	DOI= "10.1051/0004-6361/201936385",
	url= "https://doi.org/10.1051/0004-6361/201936385",
	journal = {A\&A},
	year = 2021,
	volume = 651,
	pages = "A53",
}

@INPROCEEDINGS{Lada_1987,
       author = {{Lada}, Charles J.},
        title = "{Star formation: from OB associations to protostars.}",
    booktitle = {Star Forming Regions},
         year = 1987,
       editor = {{Peimbert}, Manuel and {Jugaku}, Jun},
       series = {IAU Symposium},
       volume = {115},
        month = jan,
        pages = {1},
       adsurl = {https://ui.adsabs.harvard.edu/abs/1987IAUS..115....1L}
}

@ARTICLE{Greene_1994,
       author = {{Greene}, Thomas P. and {Wilking}, Bruce A. and {Andre}, Philippe and {Young}, Erick T. and {Lada}, Charles J.},
        title = "{Further Mid-Infrared Study of the rho Ophiuchi Cloud Young Stellar Population: Luminosities and Masses of Pre--Main-Sequence Stars}",
      journal = {\apj},
         year = 1994,
        month = oct,
       volume = {434},
        pages = {614},
          doi = {10.1086/174763},
       adsurl = {https://ui.adsabs.harvard.edu/abs/1994ApJ...434..614G}
}

@article{Williams_2025,
    author = {Williams, Joe and Krijt, Sebastiaan and Bitsch, Bertram and Houge, Adrien and Bergner, Jennifer},
    title = {Locked in ice: how pebble drift and volatile entrapment can significantly impact carbon and oxygen ratios in evolving protoplanetary discs},
    journal = {Monthly Notices of the Royal Astronomical Society},
    volume = {544},
    number = {4},
    pages = {3562-3578},
    year = {2025},
    month = {12},
    issn = {0035-8711},
    doi = {10.1093/mnras/staf1839},
    url = {https://doi.org/10.1093/mnras/staf1839},
    eprint = {https://academic.oup.com/mnras/article-pdf/544/4/3562/64923400/staf1839.pdf},
}

@article{Krijt_2025,
doi = {10.3847/2041-8213/adfbe3},
url = {https://doi.org/10.3847/2041-8213/adfbe3},
year = {2025},
month = {sep},
publisher = {The American Astronomical Society},
volume = {990},
number = {2},
pages = {L72},
author = {Krijt, Sebastiaan and Banzatti, Andrea and Zhang, Ke and Pinilla, Paola and Kaeufer, Till and Bergin, Edwin A. and Salyk, Colette and Pontoppidan, Klaus and Blake, Geoffrey A. and Long, Feng and Huang, Jane and Colmenares, María José and Williams, Joe and Houge, Adrien and Narang, Mayank and Vioque, Miguel and Lambrechts, Michiel and Cleeves, L. Ilsedore and Öberg, Karin and The JDISCS Collaboration},
title = {Cosmic Cascades: How Disk Substructure Regulates the Flow of Water to Inner Planetary Systems},
journal = {The Astrophysical Journal Letters}
}

@article{Sturm_2022,
	author = {{Sturm, J. A.} and {McClure, M. K.} and {Harsono, D.} and {Facchini, S.} and {Long, F.} and {Kama, M.} and {Bergin, E. A.} and {van Dishoeck, E. F.}},
	title = {Tracing pebble drift and trapping using radial carbon depletion profiles in protoplanetary disks},
	DOI= "10.1051/0004-6361/202141860",
	url= "https://doi.org/10.1051/0004-6361/202141860",
	journal = {A\&A},
	year = 2022,
	volume = 660,
	pages = "A126",
}

@article{Megeath_2012,
  author  = {Megeath, S.~T. and Gutermuth, R. and Muzerolle, J. and
             Kryukova, E. and Flaherty, K. and Hora, J.~L. and
             Allen, L.~E. and Hartmann, L. and Myers, P.~C. and
             Pipher, J.~L. and Stauffer, J. and Young, E.~T. and
             Fazio, G.~G.},
  title   = {The {Spitzer} Space Telescope Survey of the {Orion} {A}
             and {B} Molecular Clouds. {I}. {A} Census of Dusty Young
             Stellar Objects and a Study of Their Mid-infrared Variability},
  journal = {\aj},
  year    = {2012},
  volume  = {144},
  pages   = {192},
  doi     = {10.1088/0004-6256/144/6/192}
}

@article{Dunham_2015,
  author  = {Dunham, M.~M. and Allen, L.~E. and Evans, N.~J. and
             Broekhoven-Fiene, H. and Cieza, L.~A. and
             Di Francesco, J. and Gutermuth, R.~A. and Harvey, P.~M. and
             Hatchell, J. and Heiderman, A. and Huard, T.~L. and
             Johnstone, D. and Kirk, J.~M. and Matthews, B.~C. and
             Miller, J.~F. and Peterson, D.~E. and Young, K.~E.},
  title   = {Young Stellar Objects in the {Gould Belt}},
  journal = {\apjs},
  year    = {2015},
  volume  = {220},
  pages   = {11},
  doi     = {10.1088/0067-0049/220/1/11}
}

@article{Whittet_2001,
doi = {10.1086/318421},
url = {https://doi.org/10.1086/318421},
year = {2001},
month = {feb},
publisher = {},
volume = {547},
number = {2},
pages = {872},
author = {Whittet, D. C. B. and Gerakines, P. A. and Hough, J. H. and Shenoy, S. S.},
title = {Interstellar Extinction and Polarization in the Taurus Dark Clouds:
The Optical Properties of Dust near the Diffuse/Dense Cloud
Interface},
journal = {The Astrophysical Journal}
}

@article{Doppmann_2005,
doi = {10.1086/431954},
url = {https://doi.org/10.1086/431954},
year = {2005},
month = {sep},
publisher = {},
volume = {130},
number = {3},
pages = {1145},
author = {Doppmann, G. W. and Greene, T. P. and Covey, K. R. and Lada, C. J.},
title = {The Physical Natures of Class I and Flat-Spectrum Protostellar Photospheres: A Near-Infrared Spectroscopic Study*},
journal = {The Astronomical Journal}
}

@misc{SPHEREx_QR1,
  doi = {10.26131/IRSA629},
  url = {https://catcopy.ipac.caltech.edu/dois/doi.php?id=10.26131/IRSA629},
  author = {{SPHEREx Team}},
  title = {SPHEREx Quick Release Spectral Images},
  publisher = {IPAC},
  year = {2025}
}

\end{document}